\documentclass[
aps,                    
prd,                    
superscriptaddress,    
nofootinbib,            
amsmath,               %
amssymb,               %
a4paper,                
floatfix]               
{revtex4}               

\usepackage{booktabs}
\usepackage{multirow}
\usepackage[colorlinks = true,
            linkcolor = blue,
            urlcolor  = blue,
            citecolor = blue,
            anchorcolor = blue]{hyperref}

\usepackage{float}
\usepackage{amssymb}
\usepackage{graphics}
\usepackage{graphicx}
\usepackage{amsmath}
\usepackage{hyperref}
\usepackage{amsfonts}
\usepackage{subfigure}
\usepackage{xcolor}
\usepackage{cancel}
\usepackage{colordvi}
\usepackage{lipsum}

\newcommand{\fr}{\frac}

\newcommand{\al}{\alpha}

\newcommand{\la}{\lambda}

\newcommand{\si}{\sigma}

\newcommand{\ga}{\gamma}

\newcommand{\de}{\delta}

\newcommand{\lcdm}{$\Lambda$CDM}
\newcommand{\gdcdm}{$\gamma\delta$CDM}

\newcommand{\be}{\begin{equation}}
\newcommand{\ee}{\end{equation}}
\newcommand{\beqa}{\begin{eqnarray}}
\newcommand{\eeqa}{\end{eqnarray}}
\newcommand{\bi}{\begin{itemize}}
\newcommand{\ei}{\end{itemize}}
\newcommand{\ben}{\begin{enumerate}}
\newcommand{\een}{\end{enumerate}}
\newcommand{\bw}{\begin{widetext}}
\newcommand{\ew}{\end{widetext}}

\begin{document}

\title{Testing \gdcdm\ Model in the Redshift Bins}

\author{Furkan \c{S}akir Dilsiz}
\email[Corresponding author: ]{furkansakirdilsiz@gmail.com}
\affiliation{Department of Physics, Mimar Sinan Fine Arts University, Bomonti 34380, \.{I}stanbul, T\"urkiye}

\author{Cemsinan Deliduman}
\email{cdeliduman@gmail.com}
\affiliation{Department of Physics, Mimar Sinan Fine Arts University, Bomonti 34380, \.{I}stanbul, T\"urkiye}

\author{Selinay Sude Binici}
\email{binici.sude@gmail.com}
\affiliation{Department of Physics, \.{I}stanbul Technical University, Maslak 34469, \.{I}stanbul, T\"urkiye}

\begin{abstract}
The Hubble crisis is the discrepancy in the values of the Hubble constant inferred from diverse observations in the late and early Universe, being of the order 5$\si$.
Instead of resolution, the conflict is getting larger with further late-time observations.
A fundamental constant should be and remain constant throughout the cosmological history and thus at all redshifts. The fact that it turns out to be a function of redshift in the \lcdm\ model points out that either there is a problem with the current cosmological model, indicating unknown new physics, or there are unknown systematics in some of the observations. In this work, we investigate the redshift dependence of the Hubble constant in the \gdcdm\ cosmological model, which is a new cosmological model based on $f(R)$ gravity in an anisotropic background. Through data analysis with the Pantheon+ type Ia supernovae, the cosmic chronometers Hubble, and both the old and the Dark Energy Spectroscopic Instrument (DESI) baryon acoustic oscillation data, we establish that the Hubble constant in our model does not evolve with redshift. We also confirm that our model fits the aforementioned data better than the \lcdm\ model by checking various information criteria. The value of the Hubble constant obtained in the \gdcdm\ model is in the 1$\si$ bound of the late Universe observations.
\end{abstract}

\maketitle

\section{Introduction \label{intro}}

Hubble constant is arguably the most important parameter of the current cosmological paradigm, namely the \lcdm\ model. As its name suggests, it is a constant and therefore its inferred value from cosmological observations, whether these come from the late Universe or the early Universe, should agree with each other inside the observational error bounds. However, whereas the inferred value of the Hubble constant from the late-time observations utilizing the distance ladder is $H_0=73.30\pm 1.04\ km/s/Mpc$ \cite{Riess:2021jrx}, in contrast it is $H_0=67.37\pm 0.54\ km/s/Mpc$ \cite{Planck:2018vyg} from the early Universe cosmic microwave background (CMB) observations. Both of these measurements are historically very precise and the fact that they differ from each other by more than 5$\si$ \cite{Riess:2021jrx,Freedman:2021ahq} is termed the Hubble crisis \cite{Verde:2019ivm}, or in a mild manner, the Hubble tension.

Hubble tension is one of the most important problems in cosmology \cite{Abdalla:2022yfr,Shah:2021onj,Pesce:2020xfe,Freedman:2019jwv,Wong:2019kwg,Birrer:2018vtm,Dominguez:2019jqc,LIGOScientific:2017adf,Moresco:2017hwt,DiValentino:2021izs,Perivolaropoulos:2021jda,Schoneberg:2021qvd,Knox:2019rjx} and with the latest observations it seems to get even worse. The James Webb Space Telescope (JWST)'s latest observations ruled out the so-called ``crowding problem'' \cite{Riess:2024ohe,Anand:2024nim,Li:2024yoe,Freedman:2023jcz}, which is the excess brightness in the observations of Cepheid variables, as a possible contributor to systematic errors in the SH0ES result \cite{Riess:2021jrx}. 
Several new studies support a value of the Hubble constant even larger than the one obtained through the calibration of the distance ladder: the most precise distance measurement to the Coma cluster and the DESI fundamental plane relation \cite{Said:2024pwm} indicates $H_0=76.5\pm 2.2\ km/s/Mpc$ \cite{Scolnic:2024hbh}; removal of the problematic data in the calibration of the Tully-Fisher relation brings about a similarly high value of $H_0=76.3\pm 2.1(stat)\pm 1.5(sys)\ km/s/Mpc$ \cite{Scolnic:2024oth}, whereas the calibration with all available data indicated $H_0=73.3\pm 2.1(stat)\pm 3.5(sys)\ km/s/Mpc$ \cite{Boubel:2024cqw}; the measurement of $H_0$ from JWST's observation of multiply imaged type Ia supernova ``SN H0pe'' produces $H_0=75.4^{+8.1}_{-5.5}\ km/s/Mpc$ \cite{Pascale:2024qjr}; and spectral modeling of type II supernova allows a determination of $H_0=74.9\pm 1.9 (stat)\ km/s/Mpc$, a result free from the distance ladder calibration \cite{Vogl:2024bum}. 
There are also some studies which indicate a value for the Hubble constant in agreement with the SH0ES result \cite{Riess:2021jrx}: according to \cite{Riess:2024vfa} JWST observations validate the Hubble Space Telescope (HST)'s distance measurements, and they obtain $H_0=72.6\pm 2.0\ km/s/Mpc$; similarly, focusing only on the first two rungs of the distance ladder indicates a value of $H_0=73.4\pm 1.0\ km/s/Mpc$ according to \cite{Huang:2024gfw}; in contrast, a renewed Cepheid variable calibration provides $H_0=72.35\pm 0.91\ km/s/Mpc$ \cite{Hogas:2024qlt}, reducing the Hubble tension. 
There are also studies which agree with the value of Hubble constant inferred from the CMB observations \cite{Planck:2018vyg}: combining Cepheids, TRGB stars, and J-region asymptotic giant branch stars data from JWST observations in \cite{Freedman:2024eph} it is determined that $H_0=69.96\pm 1.05 (stat)\pm 1.12 (sys)\ km/s/Mpc$; further, an analysis \cite{Pang:2024wul} done with non-Planck (WMAP) together with DESI baryon acoustic oscillations (BAO) data results in $H_0=68.86\pm 0.68\ km/s/Mpc$, reducing the Hubble tension; similarly, the newest calculation form DESI BAO data, that avoids the need of calibrating the value of the sound horizon, is reported \cite{Guo:2024pkx} to result in a value of $H_0=68.4^{+1.0}_{-0.8}\ km/s/Mpc$; lastly, a recalibration of DESI BAO data by utilizing the ``deep learning techniques'' \cite{Shah:2024gfu} reduced the Hubble tension further with an inferred value of the Hubble constant $H_0=71.20\pm 0.47\ km/s/Mpc$.

The Hubble tension may point out the existence of a different cosmological model or new physics. It is also plausible that the standard cosmological model, which prefers the Hubble constant less than $70\ km/s/Mpc$ according to the data, is the correct description of the Universe, and all the late-time measurements of the Hubble constant are plagued by systematic errors. Another possibility is that the Hubble constant might be ``running with redshift'' and thus it has different inferred values depending on the observations done at different redshifts. As discussed in the following, this third possibility was seen as a likely solution of the Hubble tension in several works in the literature.

However, in the Friedmann equation framework, the Hubble constant $H_0$ is a constant, and thus it should have the same value from whichever observation or the dataset it is calculated. By integrating the Friedmann equations, the Hubble constant can be correlated with the Hubble parameter as
\be \label{H0}
H_0 = H(z) \exp\Big( -\fr32 \int\limits^z_0 \fr{1+\omega_{model}(z^\prime)}{1+z^\prime} dz^\prime \Big) ,
\ee
where $\omega_{model}(z)$ depends on the cosmological model, which could contain several species of energy densities \cite{Krishnan:2020vaf,Krishnan:2022fzz,Jia:2022ycc}. Therefore, for $H_0$ to have the same value at all redshifts, Hubble parameter values inferred from observations at distinct redshifts should balance out the contribution from the exponentiated integral, which is very much model dependent due to the existence of the redshift dependent function $\omega_{model}(z)$. If this balancing act is observed not to be performed for a cosmological model, this would indicate the existence of serious problems with that model \cite{Krishnan:2020vaf,Krishnan:2022fzz,Vagnozzi:2023nrq}. 
This requirement gives us an excellent test of possible cosmological models in the Friedmann equation framework \cite{Krishnan:2022fzz}. If the Hubble parameter and the exponentiated integral do not balance out each other for a particular dataset, then observations at distinct redshifts would infer distinct values for the Hubble constant. 
This kind of result will point out that the cosmological model which specifies $\omega_{model}(z)$ cannot be correct \cite{Krishnan:2022fzz}.
There have been several analyses in the literature, based on the \lcdm\ model, that show inferred values of the Hubble constant depending on the redshift of the observation(s) from which it is inferred.
One unifying feature of these analyses is that the Hubble constant decreases with increasing redshift in various datasets. 

One of the first such analyses was performed by the H0LiCOW collaboration \cite{Wong:2019kwg,Millon:2019slk}. They determined $H_0$ from measured time delays of six lensed quasars. Even though joint analysis indicated a value of $H_0$ in agreement with the local measurements of $H_0$, individual lenses at increasing redshifts indicated a descending trend with a falling slope of $1.9\sigma$ statistical significance in the inferred values of $H_0$ \cite{Wong:2019kwg}. 
Such a descending trend was also observed when the low redshift Pantheon+ Type Ia Supernovae (Sne Ia) data, combined with BAO data, cosmic chronometers (CC) Hubble data and megamasers data, is separated into six bins according to their redshift \cite{Krishnan:2020obg}. The Hubble constant in each bin is then obtained by fitting the \lcdm\ model. Descending trend with a line of falling slope is found to have $2.1\sigma$ statistical significance.
Several groups \cite{Jia:2022ycc,Dainotti:2021pqg,Schiavone:2022shz,Dainotti:2022bzg,Dainotti:2022rea,Colgain:2022rxy,Colgain:2022tql,Malekjani:2023ple,Akarsu:2024qiq,Jia:2024wix,Lopez-Henandez:2024ekv} have further shown that there is such a descending trend in the value of the Hubble constant obtained by fitting the \lcdm\ model in several redshift bins of Pantheon+ Type Ia Supernovae (Sne Ia) data, combined with various other cosmological datasets. 
In the series of papers \cite{Dainotti:2021pqg,Schiavone:2022shz,Dainotti:2022bzg,Dainotti:2022rea}, datasets are enlarged in each of the consecutive analyses. The binned analysis done first with Pantheon SNe Ia dataset \cite{Dainotti:2021pqg,Schiavone:2022shz}, then combining the BAO data \cite{Dainotti:2022bzg}, and lastly also combining the so-called platinum sample of Gamma Ray Bursts (GRBs) \cite{Dainotti:2022rea} all pointed out a decreasing trend of the Hubble constant with redshift both in the \lcdm\ model and also the $\omega_0\omega_a$CDM model \cite{Chevallier:2000qy,Linder:2002et}. As shown in \cite{Dainotti:2021pqg,Schiavone:2022shz,Dainotti:2022bzg}, this decreasing trend is more pronounced in the $\omega_0\omega_a$CDM model compared to the \lcdm . It is also commented \cite{Schiavone:2022shz,Dainotti:2022bzg} that a possible explanation of such a decreasing trend could be found in an $f(R)$ modified theory of gravity. Similar binned analyses were also done in \cite{Colgain:2022rxy,Colgain:2022tql,Malekjani:2023ple} with the observation of a similar decreasing trend in $H_0$ and an increasing trend in the matter density, $\Omega_{m0}$ parameter. The binned analysis is first done with the Pantheon SNe Ia data, together with CC and BAO datasets \cite{Colgain:2022rxy}, then with mock Hubble data the \lcdm\ model is tested in redshift bins and the similar decreasing trend of $H_0$ is again observed \cite{Colgain:2022tql}. This analysis is repeated with the newer Pantheon+ SNe Ia dataset in \cite{Malekjani:2023ple} and possibility of a negative dark energy density in the \lcdm\ model is investigated. At the end of their analysis, the authors reached the conclusion that either there exist some unknown systematic errors in SNe Ia data or the \lcdm\ model is broken down. 

Jia, Hu and Wang \cite{Jia:2022ycc,Jia:2024wix} improved on the previous work by utilizing a piecewise redshift dependent function $H_{0,z}$, which is constant in each redshift bin. Their improvement was to eliminate correlations between different bins by diagonalizing the covariance matrix with a principal component analysis \cite{Huterer:2004ch}. In their first analysis \cite{Jia:2022ycc}, they used the Pantheon+ SNe Ia dataset together with the CC Hubble dataset and 12 BAO data points from the extended Baryon Oscillation Spectroscopic Survey (eBOSS). They found a decreasing trend in the Hubble constant with a significance of 5.6$\sigma$. Later they repeated \cite{Jia:2024wix} their analysis with the newly announced DESI BAO dataset and found a decreasing trend in the Hubble constant with an even larger 8.6$\sigma$ statistical significance. 
Another work that used DESI BAO data combined with other data sets \cite{Lopez-Henandez:2024ekv} reached to similar conclusions. More recently, the constrained phase space including the eBOSS BAO data at various redshifts is used to observe redshift-dependent trends in the parameters of the \lcdm\ model \cite{Mukherjee:2024pcg}.
These results, combined with the analyses done by other groups as described above, strongly suggest the breakdown of the \lcdm\ model.

In the present work we are going to test the recently proposed \gdcdm\ model \cite{Deliduman:2023caa,Binici:2024smk} in the redshift bins of the late-time cosmological data to attest whether it has the same pathology of running $H_0$ as in the \lcdm\ and the $\omega_0\omega_a$CDM models. If that is the case, then the \gdcdm\ model should also be declared problematic, at the least.
The \gdcdm\ cosmological model is based on $f(R)$ gravity in an anisotropic background. 
In this model, the expansion of the Universe depends very differently on the energy content of the Universe compared to the \lcdm\ model. The differences come about twofold:
Firstly, the contribution of each energy density to the Hubble parameter is weighted by an equation of state parameter dependent constant. Additionally, in their contribution to the Hubble parameter, the dependence of energy densities on redshift differs from what their physical nature requires. This change in the relation of the Universe's energy content to the Hubble parameter modifies the relation between redshift and cosmic time. It is shown \cite{Deliduman:2023caa,Binici:2024smk} that this model does not allow a cosmological constant component. However, a dynamical dark energy, decreasing with the expansion of the universe, is a possibility.
In the following sections, through data analysis with the Pantheon+ type Ia supernovae, the cosmic chronometers, and both the old and the Dark Energy Spectroscopic Instrument (DESI) baryon acoustic oscillation data, we are going to demonstrate that the Hubble constant in our model does not evolve with redshift. We will also confirm that our model fits the aforementioned data better than the \lcdm\ model by checking various information criteria. Another interesting result will be that the value of the Hubble constant obtained through this analysis is in the 1$\si$ bound of the SH0ES result \cite{Riess:2021jrx}.

This article's goal is to test the \gdcdm\ model using redshift bins of data to see if the Hubble constant runs with the redshift or maintains its value at different redshifts. This is a crucial diagnostic to test a cosmological model \cite{Krishnan:2022fzz}: redshift dependent Hubble constant is a reliable indicator of a cosmological model's breakdown. We are not attempting to determine if an evolving $H_0$ can be explained by the \gdcdm\ model. There can be no plausible justification for evolution of a model's parameter, that is constant by definition. The fact that $H_0$ runs with redshift in the \lcdm\ model indicates that --in its current form-- the \lcdm\ model is failing to accommodate the data. $H_0$ diagnostic \cite{Krishnan:2022fzz} shows that the \gdcdm\ model is a serious alternative.

There were previous works that aimed to find a theoretical framework for the running of $H_0$. In \cite{Dainotti:2021pqg,Schiavone:2022wvq} $f(R)$ gravity is expressed as a scalar-tensor theory, and the decay of the Hubble constant with redshift is attained with the dynamics of the scalar field in the theory. The parameters $\eta$ or $\al$ in their $H_0^{eff}$ \cite{Dainotti:2021pqg,Schiavone:2022wvq} are related to the parameter $\de$ in the \gdcdm\ model, however, the similarity stops there. Their model is very different from ours and supports the decay of $H_0$ with redshift. Similar to our redshift dependent dark energy, an evolutionary dark energy model was proposed in \cite{Montani:2024ntj}, where a binned analysis of the Pantheon SNe Ia sample was performed, and it is shown that this model also supports a running Hubble constant. 
In some scalar-tensor theories \cite{Boisseau:2010pd,Akarsu:2019pvi} solutions of the field equations resulted in the dependence of the energy densities on redshift similar to the \gdcdm\ model. These solutions are not exact, unlike the \gdcdm\ model, and dark energy can only be included effectively. Moreover, in these works, the contribution of dust to the expansion diminishes faster than its actual physical dynamics demands, contrary to the \gdcdm\ model.

This paper is organized as follows: In the next section, the derivation of the \gdcdm\ model in an anisotropic background in the $f(R)$ gravity framework is summarized. In Section \ref{data} we present datasets with which we perform the Bayesian analysis, which is summarized in Section \ref{method}. In section \ref{results} we present and discuss the main results of our analyses. Lastly, in Section \ref{conc}, we summarize our work and examine our results.

\section{Theory \label{theory}} 

In this section, we summarize the fundamentals of the $f(R)$ modified gravity in the Bianchi type I background and review the resulting \gdcdm\ cosmological model. Further details can be found in \cite{Deliduman:2023caa,Binici:2024smk}.

The action of $f(R)$ gravity is given in terms of an arbitrary function of the scalar curvature $R$. Together with the matter part, the total action can be written as
\be \label{action}
S=-\int d^4 x \sqrt{-g}\left( \fr1{2\kappa} f(R) + \mathcal{L}_m \right) ,
\ee
where $\mathcal{L}_m$ is the Lagrangian density for matter fields and the Einstein constant is denoted by $\kappa = 8\pi G$. We use natural units (c=1) throughout this article.
Varying the action (\ref{action}) with respect to the metric, the field equations are found to be
\be \label{field}
f_R R_{\mu\nu} -\fr12 f g_{\mu\nu} +\left( g_{\mu\nu}\Box - \nabla_\mu \nabla_\nu \right) f_R = \kappa T_{\mu\nu} ,
\ee
where the energy-momentum tensor is obtained from $T_{\mu\nu} = -\fr{2}{\sqrt{-g}} \fr{\delta \mathcal{S}_m}{\delta g_{\mu\nu}}$, and $f_R$ denotes $\fr{\partial f}{\partial R}$.

Based on various observational indications \cite{Schwarz:2004gk,Land:2005ad,Copi:2013jna,Buchert:2015wwr,Schwarz:2015cma,Planck:2019kim,Planck:2019evm} and theoretical works \cite{Campanelli:2006vb,Campanelli:2007qn,Cea:2019gnu,Luongo:2021nqh,Krishnan:2021jmh,Cea:2022mtf} we hypothesize the background metric to be anisotropic. The simplest case of such an anisotropic background is the Bianchi type I spacetime, given by the metric
\be \label{B1}
ds^2=-dt^2+A(t)^2dx^2+B(t)^2 (dy^2+dz^2) \, ,
\ee
where $A(t)$ and $B(t)$ are directional scale parameters, distinct from each other. Then we define the directional Hubble parameters as $H_x = \dot{A}/A$ and $H_y = H_z = \dot{B}/B$. The average scale parameter, $a(t)$, is defined in terms of the physical volume as $V(t) = A\cdot B^2 = a(t)^3$. In terms of the directional Hubble parameters we also define average quantities, which are the average Hubble parameter and the shear anisotropy parameter, respectively:
\be
H (t) = \fr13 ( \frac{\dot{A}}{A} +2\frac{\dot{B}}{B} ) \quad \mathrm{and} \quad S (t) =\frac{\dot{A}}{A} - \frac{\dot{B}}{B} \, .
\ee
Note that $S^2(t)$ quantifies the anisotropic expansion and is related to the shear scalar $\si^2(t)$ \cite{Leach2006} by $S^2(t) = 3\si^2(t)$.
The average Hubble parameter can be further expressed in terms of the average scale parameter as $H(t) = \dot{a}/a$.

Expressing the field equations in terms of the parameters $H(t)$ and $S(t)$, it was found \cite{Deliduman:2023caa,Binici:2024smk} that the field equations depend on the time derivative of the shear anisotropy parameter. We then made the basic assumption that the energy momentum tensor is in the form of a sum of multiple perfect fluid components. Since the background is anisotropic, this assumption requires 
the trace free Gauss--Codazzi equation \cite{Leach2006,Banik2016} given by
\be \label{GC}
\fr{\dot{S}}S = - \left[ 3 \fr{\dot{a}}a + \fr{\dot{f}_R}{f_R} \right]
\ee
to hold, so that the field equations, written in terms of $H(t)$ and $S(t)$,
\beqa 
\label{iso} \kappa \rho &=& - f_R \left( 3 H^2+3 \dot{H} +\frac{2 S^2}{3} \right) + \fr12 f + 3H \dot{f}_R \, ,  \\
\label{iso1} \kappa p &=&  f_R \left( 3 H^2+\dot{H} \right) - \fr12 f - 2H \dot{f}_R - \ddot{f}_R \, ,
\eeqa 
remain self-consistent after the perfect fluid assumption. Equation (\ref{GC}) can be solved after making the choice $\dot{S} = -(3 - \delta) HS$, which is inspired by the behavior of the shear scalar $\si^2(t)$ in \cite{Leach2006}, where the dynamical analysis limits $\delta$ to $0 < \delta < 1$. Then $S(t)$ and $f_R(t)$ are solved to have the forms $S(z) = s_0 (z+1)^{(3 - \delta)}$ and $f_R (z) = \varphi (z+1)^\delta$, respectively, in terms of redshift, $z+1 = 1/a$. Here, $\varphi$ and $s_0$ are integration constants. 

If the expansion of the Universe is due to the combined effect of the multiple perfect fluid components, the expected general form of the average Hubble parameter is a polynomial function of redshift, $z$, given by
\be \label{H2a}
H^2 (z) = \sum_\la h_\la (z+1)^\la \, ,
\ee
where the possible values of $\la \in \mathbb{R}$ are to be fixed by the field equations (\ref{iso},\ref{iso1}).

The perfect fluid components include dust (m) with vanishing pressure, radiation (r) with positive pressure, and
dark energy (e) with negative pressure. It is argued in \cite{Deliduman:2023caa,Binici:2024smk} that a cosmological constant component cannot be a part of this theory. The equation of state parameter of dark energy component is given by $\omega = \gamma/3 -1$ with $0< \gamma \le 2$. Thus, the total energy density is given by
\be \label{emt}
\rho (z) = \rho_{e0} (z+1)^\gamma + \rho_{m0} (z+1)^3 + \rho_{r0} (z+1)^4  \, ,
\ee
where $\rho_{e0}, \rho_{m0}$ and $\rho_{r0}$ denote the present-day dark energy, dust and radiation densities, respectively.

Solving the field equations (\ref{iso},\ref{iso1}) with the given forms of $H(z)$ (\ref{H2a}) and $\rho (z)$ (\ref{emt}) we are able to determine the Hubble parameter in terms of the dimensionless density parameters as
\be \label{H2}
H^2 = \fr{H_0^2/\varphi}{(z+1)^\delta} \left[ \fr{\Omega_{e0}}{b_\gamma} (z+1)^{\gamma} 
+ \fr{\Omega_{m0}}{b_3} (z+1)^{3} + \fr{\Omega_{r0}}{b_4} (z+1)^{4} 
+ \fr{\Omega_{s0}}{1 -\delta} (z+1)^{6 - \delta} \right] 
\quad \mathrm{with} \quad \Omega_{s0} = \fr{\varphi s_0^2}{9 H_0^2} \, .
\ee
where $\Omega_{e0}, \Omega_{m0}$ and $\Omega_{r0}$ denote the present dimensionless density parameters for dark energy, dust and radiation components, respectively. They are obtained by dividing respective densities with the critical density $\rho_c = 3H_0^2/8\pi G$.
Additionally, $\Omega_{s0}$ denotes the present-day dimensionless contribution of anisotropic shear to the Hubble parameter. $H_0$ is the Hubble constant at the present time.
The coefficients $b_n$ ($n = \gamma,3,4$) in (\ref{H2}) are given by
\be \label{bn}
b_n = - \left( 1+ \delta - \fr1{2n} (n -\delta)(4+\delta) \right) .
\ee
Since the coefficient $b_0$ is divergent unless $\delta = 0$, this model does not allow inclusion of a cosmological constant term. Dark energy term need to be redshift dependent. 

The general relativistic case is the limit $\delta \rightarrow 0$ and $\varphi \rightarrow 1$. In this limit, one finds $b_\gamma = b_3 = b_4 = 1$ and $\Omega_{e0} + \Omega_{m0} + \Omega_{r0} + \Omega_{s0} = 1$ as is expected for a flat universe described by the general relativistic Friedmann equations. In the same limit, shear dissipates with the sixth power of $a(t)$ as is the case in the anisotropic \lcdm\ model \cite{Akarsu:2019pwn}.
To simplify the model (\ref{H2}) we will also set $\varphi = 1$ in the general $f(R)$ case when we constrain the cosmological parameters with Bayesian data analysis. 
A non-vanishing value of the $\delta$ parameter affects both how the different perfect fluid components contribute to the expansion of the Universe and also how these components dissipate as the Universe expands. The coefficients $b_n$ for $n = \gamma,3,4$ are the weight factors and affect how different $\Omega_{i0}$ contribute to the Hubble parameter. Depending on the value of $\delta$, the contribution of one component may get enhanced as the contribution of another component diminishes. 
A nonzero value of $\delta$ also changes how the contribution of perfect fluid components to the expansion of the Universe diminishes with the expansion. For example, even though the dust component physically diminishes with $a^3$, its contribution to the expansion diminishes slower with $a^{3-\delta}$. This is also the case for the relativistic perfect fluid components. The redshift dependence of their contributions to the Hubble parameter has lower exponents compared to what their actual physical dynamics requires.

\section{Data \label{data}} 

The datasets which we use to constrain the cosmological parameters of the \gdcdm\ and the \lcdm\ cosmological models are the same datasets that are used by Jia, Hu and Wang in \cite{Jia:2022ycc,Jia:2024wix}: these  are the Pantheon+ SNeIa, the CC Hubble, the older eBOSS BAO and the newer DESI BAO datasets.

\subsection{Pantheon+ SNe Ia data}

We use the Pantheon+ sample of \cite{Brout:2022vxf,Scolnic:2021amr}. The Pantheon+ sample contains 1701 light curves of 1550 distinct SNe Ia in the redshift range $0.001 < z < 2.262$ . The data sample can be found in the GitHub \cite{Scolnic:pant}. 
For the peculiar velocities of supernovae with low redshifts not to impact our analysis, we do not include data with redshift $z<0.01$. Thus, we use 1590 data points in the redshift range $0.01 < z < 2.26$.

The distance modulus residuals are given by
\be
\Delta \mu_i = \mu_{\text{dat},i}-\mu_{\text{th}}(z_i)  ,
\ee
where the observed distance modulus $\mu_{\text{dat},i}$ can be found in the Pantheon+ data file. Whereas the model-dependent distance modulus is defined by
\be
\mu_{\text{th}} = 5 \log_{10} \left(\fr{d_L}{1\text{Mpc}}\right) +25,
\ee
where $d_L$ is the luminosity distance, the definition of which depends on both the observational and the model parameters \cite{Alonso-Lopez:2023hkx} as
\be
d_L  = c \left(1+z_{hel}\right) \int^{ z_{HD}} _0 \fr {dz'}{H(z')},
\ee
where $z_{hel}$ is the heliocentric redshift and $z_{HD}$ is the Hubble diagram redshift. These data can be read from the  \texttt{`zhel'} and \texttt{`zHD'} columns of the Pantheon+ data file \cite{Scolnic:pant}.

The chi-squared function for the Pantheon+ sample is given by
\be
\chi^2 _{\text{Pan+}} = \Delta \mu ^T C_{Pan+}^{-1} \Delta \mu ,
\ee
where the covariance matrix $C_{Pan+}$ contains both statistical and systematic errors.

\subsection{CC Hubble data}

The observational Hubble parameter data are obtained using the cosmic chronometers technique \cite{Jimenez:2001gg}.  We use 32 Hubble parameter measurements in the redshift range $0.07\le z \le 1.965$ \cite{Yu:2017iju,Cao:2022ugh}. These data are given in Table 1 of \cite{Deliduman:2023caa} with related references \cite{Jimenez:2003iv,Simon:2004tf,Stern:2009ep,Zhang:2012mp,Moresco:2012jh,Moresco:2015cya,Moresco:2016mzx,Ratsimbazafy:2017vga,Borghi:2021rft}. 
 
The chi-squared function for the 32 H(z) measurements, denoted by $\chi^2_{CC}$, is given in terms of the covariance matrix by
\be
\chi^2_{CC} = M^T C_{CC}^{-1}M,
\ee
where $C_{CC}$ is the covariance matrix for the cosmic chronometers data, and M is the residual between the model prediction $H^{th}(z_i)$ and the observational data $H^{obs}(z_i)$ given by 
\be
M_i = H^{obs}(z_i)-H^{th}(z_i)\ .
\ee 

The uncertainties resulting from the stellar library, the stellar population synthesis model, the IMF, and the star formation history all contribute to the model covariance matrix for the cosmic chronometers data. The method we use to calculate the covariance matrix and the specific details are identical to those in \cite{Deliduman:2023caa,Moresco:2020,Moresco:2024wmr}. 

The aforementioned systematic uncertainties contribute to the off diagonal elements of the model covariance matrix for the data points extracted from \cite{Moresco:2012jh,Moresco:2015cya,Moresco:2016mzx}.  The other data points in Table 1 of \cite{Deliduman:2023caa} are included diagonally in the covariance matrix since they have no correlation with the data from \cite{Moresco:2012jh,Moresco:2015cya,Moresco:2016mzx}.

\subsection{eBOSS BAO data} \label{baodata}

We use 12 baryon acoustic oscillation (BAO) measurements in the redshift range $0.122\le z \le 2.334$. BAO data points are given in Table 2 of \cite{Jia:2022ycc} with related covariance matrices and references \cite{DES:2017rfo,Carter:2018vce,eBOSS:2020lta,eBOSS:2020hur,eBOSS:2020tmo,eBOSS:2020gbb,eBOSS:2020uxp}.

The chi-squared function for the eBOSS BAO dataset is given by \cite{Jia:2022ycc}
 \be \label{sdss}
 \chi^2 _{eBOSS} =  \Delta\nu^T C_{eBOSS} ^{-1} \Delta\nu 
 \quad \mathrm{with} \quad \Delta\nu = \nu_{dat}(z_i)-\nu_{model}(z_i) ,
 \ee
where $\nu_{dat}(z_i)$ is the vector of the BAO measurements, $D_M/r_d$, $D_H/r_d$, $D_A/r_d$ and $D_V(r_{d,f}/r_d)$, at each redshift $z_i$.
In these measurements $D_M (z)$ is the comoving distance, defined by
\be
D_M (z) = c \int^z _0 \fr {dz'}{H(z')} \ ;
\ee
$D_H(z)=c/H(z)$ is the Hubble distance; $D_A(z)$ is the angular diameter distance, defined by 
\be
D_A(z)= \fr{D_M(z)}{z+1}\ ; 
\ee
and $D_V(z)$ is the angle-averaged distance, defined by
\be
D_V (z) = \left[ \fr{czD_M ^2(z)}{H(z)} \right]^{1/3} .
\ee
The comoving size of the sound horizon at the baryon drag is given by
\be
r_d(z_d) = \int ^{\infty} _{z_d} \fr{c_s(z)}{H(z)} dz,
\ee 
where $z_d=1059.6$ is redshift at the baryon drag \cite{Planck:2015fie}. The sound speed is defined as
\be
c_s=\fr{c}{\sqrt{3(1+R_b/(1+z))}} ,
\ee
where $R_b=31500\cdot \Omega_{b0}h^2(T_{CMB}/2.7)^{-4}$, with  $T_{CMB}=2.725$ is the observed average temperature of the cosmic microwave background radiation, and the baryon density is given in terms of $\Omega_{b0}h^2=0.02226$ \cite{Planck:2015fie}.

\subsection{DESI BAO data}

We use the BAO measurements from the first year of observations of DESI collaboration \cite{DESI:2024mwx}. 12 data points in the redshift range $0.295\le z \le 2.33$ are shown in Table 1 of \cite{DESI:2024mwx}. DESI BAO sample consists of the following tracers: the bright galaxy sample (BGS), the luminous red galaxy sample (LRG), the emission line galaxy sample (ELG), the quasar sample (QSO) and the Lyman-$\al$ forest sample (Lya QSO). 

The LRG, ELG and Lya QSO data are correlated \cite{Pourojaghi:2024bxa}. For these 10 data points the chi-squared function is given by
\be
\chi^2 _{corr} =  \Delta\upsilon^T C_{DESI} ^{-1} \Delta\upsilon 
 \quad \mathrm{with} \quad \Delta\upsilon = \upsilon_{dat}(z_i)-\upsilon_{model}(z_i) ,
\ee
and for the uncorrelated BGS and QSO data we have
 \be
 \chi^2 _{uncorr} = \sum _{i=1} ^{2} \fr{\left[\upsilon_{\text{dat}}(z_i)
 -\upsilon_{\text{th}}(z_i)\right]^2}{\sigma^2 _i} ,
 \ee
 where $\upsilon_{dat}(z_i)$ is the vector of the DESI BAO measurements, $D_M/r_d$, $D_H/r_d$ and $D_V/r_d$, at each redshift $z_i$. The inverse covariance matrix, $C_{DESI} ^{-1}$, for the correlated data is given in \cite{Pourojaghi:2024bxa}. The total chi-squared function for the DESI BAO dataset is then given by
 \be \label{desi}
 \chi^2 _{DESI} = \chi^2 _{corr} +\chi^2 _{uncorr} .
 \ee

\section{Methodology \label{method}}

\subsection{Piecewise function $H_{0,z}$ \label{H0z}}

To test the possible redshift dependence of the Hubble constant in a cosmological model, we assume a piecewise function of the Hubble constant, whose value depends on the observations in specific redshift ranges, as was previously done in \cite{Jia:2022ycc}. The piecewise function of the Hubble constant means that $H_0$ has constant $H_{0,z_i}$ values in each of the corresponding redshift ranges $z_{i-1} \leqslant z < z_i$, where $i$ labels each redshift range or bin.

The constant values of $H_{0,z_i}$  in each bin can be deduced from the cosmological model by Bayesian data analysis. In this article, we determine the piecewise functions $H_{0,z}$ for both the standard flat \lcdm\ and the \gdcdm\ models.

\subsubsection{\lcdm\ model \label{lcdm}}

The Hubble parameter in the \lcdm\ model is given by
\be \label{Hz}
H(z)=H_0\left[ \Omega_{e0}+\Omega_{m0}\left(1+z\right)^3+\Omega_{r0}\left(1+z\right)^4\right]^{1/2} .
\ee
Since the radiation density today, $\Omega_{r0}$, is very small $\Omega_{r0} \sim \mathcal{O}(10^{-4})$ \cite{Mukhanov:2005sc} compared to $\Omega_{e0}\sim \Omega_{m0}\sim \mathcal{O}(10^{-1})$ \cite{Planck:2018vyg}, in the case of observations in the late Universe, i.e., observations with low redshift values, we can ignore contribution of the radiation term to the Hubble parameter.

Then, in the the \lcdm\ model, the Hubble parameter can be written in an integral form as \cite{Jia:2022ycc}
\be \label{Hzi}
H(z) = H_0 \int\limits^{z}_0 \fr{3\Omega_{m0}(1+z^\prime)^2}
{2\left[ \left(1-\Omega_{m0}\right)+\Omega_{m0}(1+z^\prime)^3 \right]^{1/2}}dz^\prime + H_0 ,
\ee
where $\Omega_{e0}$ is replaced with $1-\Omega_{m0}$. This form of the Hubble parameter is argued to be more advantageous \cite{Jia:2022ycc} to infer the Hubble constant values in redshift intervals.

We now replace the Hubble constant $H_0$ in (\ref{Hzi}) with the piecewise function $H_{0,z}$ with constant values $H_{0,z_i}$ in each redshift bin, labeled by $i$, to reach an expression for the Hubble parameter given as
\be \label{Hzisum}
H(z) = \sum\limits_{i=1}^N \left[ H_{0,z_i} \int\limits^{z_i}_{z_{i-1}} \fr{3\Omega_{m0}(1+z^\prime)^2}
{2\left[ \left(1-\Omega_{m0}\right)+\Omega_{m0}(1+z^\prime)^3 \right]^{1/2}}dz^\prime \right] + H_{0,z_N} ,
\ee
where $z_0 = 0$ and $z_N = z$. This is the same form of $H(z)$ previously obtained in \cite{Jia:2022ycc}. In that work, it is shown that for constant $H_{0,z}$ up to redshift $z$, i.e., all $H_{0,z_i} = H_0$ in each redshift bin, and $\Omega_{m0} = 0.3$ fixed, equation (\ref{Hz}) and (\ref{Hzisum}) produce the same values for the Hubble parameter $H(z)$.

 Bayesian analysis, to determine $H_{0,z_i}$ in each redshift bin in the \lcdm\ model, is also done with $\Omega_{m0} = 0.3$ fixed. As will be discussed in the next section, our results completely agree with the results of \cite{Jia:2022ycc} and \cite{Jia:2024wix} in the case of the \lcdm\ model. This was done to test that our code works correctly and that we can trust using it to test whether the Hubble constant is redshift dependent also in the \gdcdm\ model.

\subsubsection{\gdcdm\ model \label{gdcdm}}

The Hubble parameter in the \gdcdm\ model is given by:
\be \label{fgd}
H(z) = H_0 \left( \fr{\Omega_{e0}}{b_\ga} \left(1+z\right)^{\ga-\de}+\fr{\Omega_{m0}}{b_3}\left(1+z\right)^{3-\de}+\fr{\Omega_{r0}}{b_4}\left(1+z\right)^{4-\de}+\fr{\Omega_{s0}}{1-\de}\left(1+z\right)^{6-2\de} \right)^{1/2} .
\ee
Since the anisotropic shear, dimensionlessly quantified by $\Omega_{s0}$, is very small $\Omega_{s0} \sim \mathcal{O}(10^{-12})$ \cite{Deliduman:2023caa} compared to $\Omega_{e0}$ and $\Omega_{m0}$ \cite{Planck:2018vyg}, we can ignore the contribution of the anisotropic shear and of the radiation term (as in the \lcdm\ model) to the Hubble parameter in the late Universe.

Then the integral form of the Hubble parameter in the \gdcdm\ model is given by
\be \label{Hzig}
H(z) = H_0 \int\limits^{z}_0 \fr{\left[ \left(1-\fr{\Omega_{m0}}{b_3}\right) (\ga-\de) (1+z^\prime)^{\ga-\de-1}
+ \fr{\Omega_{m0}}{b_3}(3-\de) (1+z^\prime)^{2-\de} \right] }
{2\left[ \left(1-\fr{\Omega_{m0}}{b_3}\right)(1+z^\prime)^{\ga-\de}
+ \fr{\Omega_{m0}}{b_3}(1+z^\prime)^{3-\de}\right]^{1/2}}dz^\prime + H_0 .
\ee

As it is done in the \lcdm\ case above, we replace the Hubble constant $H_0$ in (\ref{Hzig}) with the piecewise function $H_{0,z}$ with constant values $H_{0,z_i}$ in each redshift bin, labeled by $i$, to reach an expression for the Hubble parameter given by
\be \label{Hzigd}
H(z) = \sum\limits_{i=1}^N \left[ H_{0,z_i} \int\limits^{z_i}_{z_{i-1}} 
\fr{\left[ \left(1-\fr{\Omega_{m0}}{b_3}\right) (\ga-\de) (1+z^\prime)^{\ga-\de-1}
+ \fr{\Omega_{m0}}{b_3}(3-\de) (1+z^\prime)^{2-\de} \right] }
{2\left[ \left(1-\fr{\Omega_{m0}}{b_3}\right)(1+z^\prime)^{\ga-\de}
+ \fr{\Omega_{m0}}{b_3}(1+z^\prime)^{3-\de}\right]^{1/2}}dz^\prime \right] + H_{0,z_N} ,
\ee
where $z_0 = 0$ and $z_N = z$. Since we are only interested in the values of $H_{0,z_i}$ in the present work, we are going to fix the other model parameters, $\Omega_{m0}$, $\ga$, and $\de$, using the complete dataset. We will then determine in each bin the values of $H_{0,z_i}$ and see if they run with redshift.

\subsection{Bayesian analysis \label{Bayes}}

We use the Bayesian inference method \cite{Padilla:2019mgi,Hogg:2010} to constrain the parameters of the \gdcdm\ and the \lcdm\ cosmological models. This method determines the maximum of the likelihood function defined as
\be \label{like}
\mathcal{L}\left(\mathcal{I}|\mathcal{\theta},\mathcal{M}\right) \propto \exp \left[ -\fr{1}{2} \chi^2 \left(\mathcal{I}|\mathcal{\theta},\mathcal{M}\right)\right]
\ee
for a model. Here $\mathcal{M}$, $\mathcal{\theta}$ and $\mathcal{I}$ represent the model, parameters of the model and the datasets, respectively. $ \chi^2 \left(\mathcal{I}|\mathcal{\theta},\mathcal{M}\right)$ is the chi-squared function. The minimum value of the chi-squared function corresponds to the maximum value of the likelihood function. The total chi-squared function is given by
\be
\chi^2 _{total} = \chi^2 _{Pan+} +\chi^2 _{CC} + \chi^2 _{BAO} ,
\ee
where $\chi^2 _{BAO}$ stands for either $\chi^2 _{eBOSS}$ (\ref{sdss}) or $\chi^2 _{DESI}$ (\ref{desi}).
We use the Markov Chain Monte Carlo (MCMC) code emcee\footnote{\url{https://emcee.readthedocs.io/en/stable/}}  \cite{Foreman-Mackey:2012any} to constrain the parameters of the cosmological models. We modified and used the code found in the Github repository \cite{Jojo2022} provided by the authors of \cite{Jia:2022ycc}.

In the case of \lcdm\ model, mass density parameter is fixed to $\Omega_{m0} = 0.3$ and uniform prior distribution $50\le H_{0,z_i} \le 80$ ($km/s/Mpc$) is used for all $z_i$. These are the same choices as were made in \cite{Jia:2022ycc}. 

In comparison, the \gdcdm\ model has two more free parameters, which are $\ga$ and $\de$. We fixed the anisotropic shear density parameter at the value of $\log_{10}(\Omega_{s0}) = -12.28$, obtained in a previous analysis of the \gdcdm\ model \cite{Deliduman:2023caa}. The radiation density parameter is given by $\Omega_{r0}=h^{-2}(2.469.10^{-5})(1+\fr78(\fr4{11})^{4/3}N_{eff})$, where $h=H_0/100$ and $N_{eff}=3.046$ \cite{Mukhanov:2005sc}.
We first run the code with the full dataset with uniform prior distributions for $H_0$, $\Omega_{m0}$, $\ga$ and $\de$ with the following ranges: $50\le H_0 \le 80$ ($km/s/Mpc$), $0.2 \le \Omega_{m0} \le 0.4$, $0.001 \le \ga \le 2.0$ and $0.0\le \de \le 0.772$. We then fixed the parameters $\Omega_{m0}$, $\ga$, and $\de$ to the median values of the posterior results. In the binned analysis we only used $H_{0,z_i}$ as the free parameter with uniform prior distribution $50\le H_{0,z_i} \le 80$ ($km/s/Mpc$) for all $z_i$. 
We verified that fixing the parameters $\Omega_m,\ \gamma$, and $\delta$ in the binned analysis does not affect the results significantly.

To specify the value of the Hubble constant, it is necessary to establish a value for the sound horizon at the baryon drag, $r_d$. Specifically, BAO data can only constrain the value of the product $r_d H_0$. Further independent cosmological observations are necessary to ascertain $r_d$ and $H_0$ separately. The observations from the CMB, analyzed within the framework of the \lcdm\ model, yield a value of $r_d =  147.1$ Mpc [2], which is also employed here in the analyses done using the \lcdm\ model. In the case of the \gdcdm\ model, we used the value of $r_d$ obtained from the model parameters to keep the analyses independent of the \lcdm\ model.

In the binned analysis, the values of the Hubble constant $H_{0,z_i}$ at different bins are correlated.
In the present work, we follow the same method described in \cite{Jia:2022ycc} to remove correlations. They used the principal component analysis method \cite{Huterer:2004ch} to remove correlations by diagonalizing the covariance matrix. We modified and used the related code generously provided by the authors of \cite{Jia:2022ycc} in the Github repository \cite{Jojo2022}.

We generate contour plots with GetDist \cite{Lewis:2019xzd}.

\section{Results and Discussion \label{results}} 

The analysis of the late-time, i.e., small redshift, cosmological data in redshift bins can be done in several distinct ways: In \cite{Krishnan:2020obg} the full dataset is divided into bins nonuniformly, aiming for weighted average redshifts of each individual dataset to coincide in each bin. In \cite{Colgain:2022rxy}, the highest redshift of each bin is kept constant, while the lowest redshift of each bin progressively gets larger. Thus, each consecutive bin contains fewer number of data points. In \cite{Malekjani:2023ple}, the SNe Ia sample is divided into two bins at a redshift $z_{split}$. As $z_{split}$ is increased, data is divided into different unequal parts each time. They observe a very pronounced decreasing trend of $H_0$ in high redshift bins with small numbers of data points. Dividing relevant data into bins with an approximately equal number of data points \cite{Dainotti:2021pqg,Schiavone:2022shz}, the so-called equal-number method, is another way to bin. Complementary to this method is to divide the data into bins of equal redshift width \cite{Huterer:2004ch}. Both of the last two ways of binning are used in the analyses performed in \cite{Jia:2022ycc,Jia:2024wix}. Since we want to compare and contrast our results with theirs, we also do our analyses with both binning methods.

In the following, we present the results of the data analysis of the \gdcdm\ and the \lcdm\ models with the CC+eBOSS+Pan+ dataset with equal-number and equal-width binning methods. Afterwards, we also present the results of the analyses with the new DESI BAO dataset replacing the older eBOSS BAO dataset. We are going to tabulate our results both in the case of the \gdcdm\ model and the \lcdm\ model. We will also present our results in graphical forms and share the 2D and 1D posterior distribution plots. The models will be compared with the information criteria in each analysis separately.

\subsection{Analysis with the CC+eBOSS+Pantheon+ dataset \label{cbp}}

The CC+eBOSS+Pantheon+ dataset includes 32 cosmic chronometers Hubble data points, 12 BAO measurement data points, and 1590 light curves of distinct Type Ia supernovae. The details of individual datasets are given in Section \ref{data}. We analyze both the \lcdm\ and the \gdcdm\ models first with the full dataset, then with the binned dataset of 8 bins with equal number data points and finally with the binned dataset of 10 bins with equal redshift width.

\subsubsection{Analysis with the full dataset \label{fcbp}} 

To keep the uncertainty in $H_{0,z}$ to be less than $1.0$, the authors of \cite{Jia:2022ycc,Jia:2024wix} fixed the dimensionless matter density parameter, $\Omega_{m0}=0.3$ and chose uniform prior distribution $50\le H_{0,z_i} \le 80$ ($km/s/Mpc$) for all $z_i$. When we constrain the flat \lcdm\ model with the full CC+eBOSS+Pan+ dataset, with the same choices, the Bayesian analysis resulted in a posterior value for the only free parameter to be $H_0 = 73.20\pm 0.12\ km/s/Mpc$. This value is close to the late-time results as described in the Introduction. However, we will discuss how much the flat \lcdm\ model is compatible with the full dataset compared to the \gdcdm\ model at the end of this subsection. In the Bayesian analysis with the BAO data one also needs the value of the sound horizon at the baryon drag. For \lcdm\ this value is taken to be $r_d =  147.1$ Mpc \cite{Planck:2018vyg}.

\begin{figure}[h]
\centering
    \includegraphics[width=0.46\textwidth]{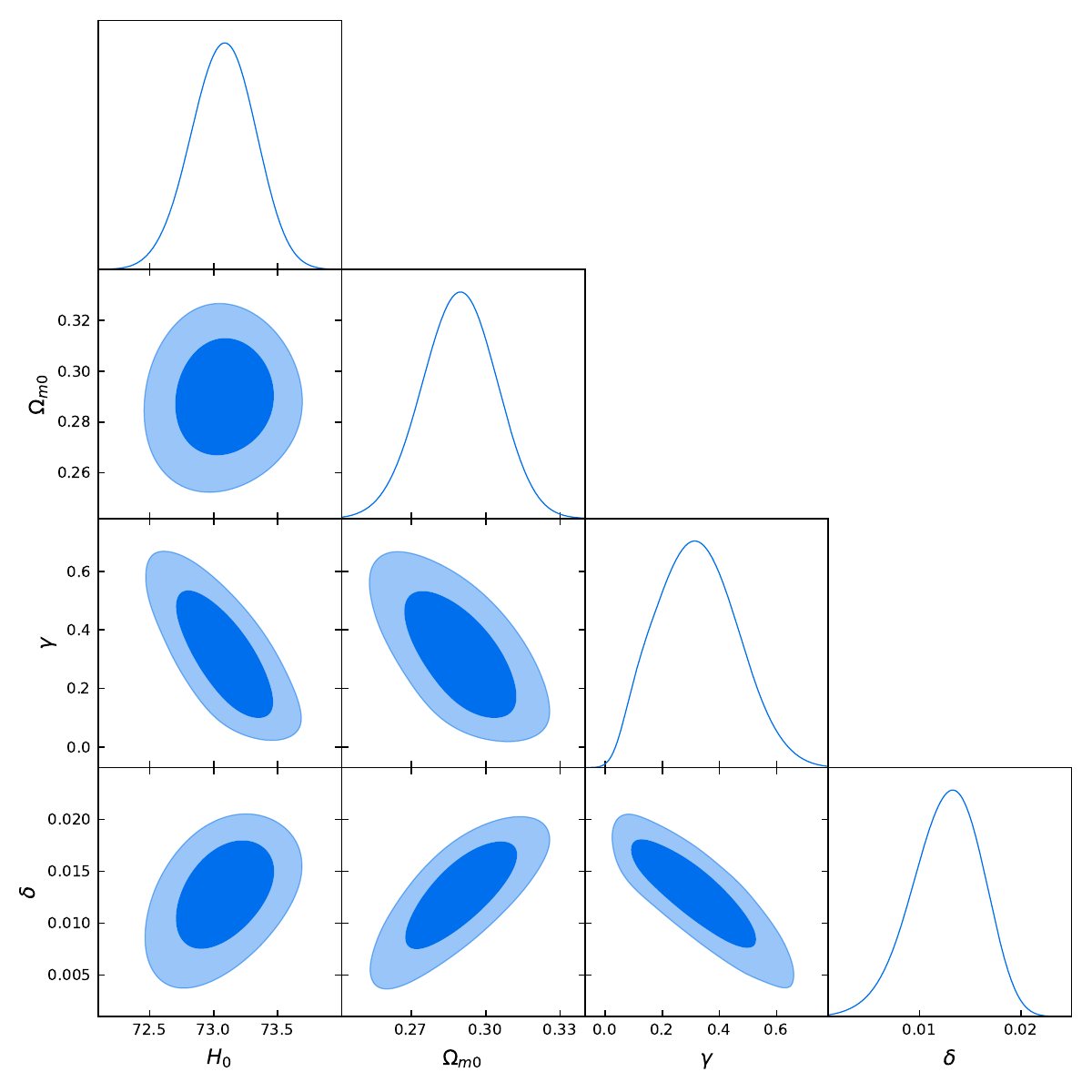}
\caption{The contour plots for 2D joint posterior distributions, with 1$\sigma$ and 2$\sigma$ confidence regions, for the free cosmological parameters ($H_0,\ \Omega_{m0},\ \gamma$ and $\delta$) of the \gdcdm\ model for the full CC+eBOSS+Pan+ dataset, together with 1D marginalized posterior distributions.}
\label{fullgdb}
\end{figure}

In the case of the full \gdcdm\ model, other than $H_0$ and $\Omega_{m0}$, there are three more free cosmological parameters; $\ga$, $\de$, and $\Omega_{s0}$. We fix the values of $\Omega_{m0}$, $\ga$, and $\de$ by constraining them with the full dataset. We then use these fixed values and leave only $H_{0,z}$ as the free parameter in the binned analyses. Due to the degeneracy between $\Omega_{s0}$ and $\de$, we first fix $\log_{10}(\Omega_{s0}) = -12.28$, obtained in a previous analysis of the \gdcdm\ model \cite{Deliduman:2023caa}. The radiation density parameter is also fixed by the Stefan-Boltzmann law and the CMB temperature \cite{Mukhanov:2005sc}. Bayesian analysis with these fixed values and uniform prior distributions for free parameters, as described in section \ref{Bayes}, resulted in the following median posterior values with $1\sigma$ confidence intervals for the cosmological parameters: $H_0 = 73.08\pm 0.25\ km/s/Mpc$, $\Omega_{m0} = 0.290\pm 0.015$, $\ga = 0.317\pm 0.15$ and $\de = 0.013\pm 0.004$. In Fig. \ref{fullgdb} we present the marginalized 1D and 2D posterior distributions of the cosmological parameters of the \gdcdm\ model.
In the Bayesian analyses with the binned datasets we fixed $\Omega_{m0}$, $\ga$ and $\de$ to these values and left only the parameters $H_{0,z_i}$ in each bin free with uniform prior distribution $50\le H_{0,z_i} \le 80$ ($km/s/Mpc$) for all $z_i$. The sound horizon in the case of the \gdcdm\ model is calculated from the values of the cosmological parameters from the full dataset fit and is found to be $r_d = 135.5$ Mpc. This value of the sound horizon is in the 1$\sigma$ range of the model-independent estimate from a recalibration of eBOSS and DESI BAO datasets, obtained by deep learning techniques \cite{Shah:2024gfu}.

\subsubsection{Binning with Equal-Number Method \label{benm}} 

\begin{figure}[h!]
\centering
\begin{subfigure}
    \centering 
    \includegraphics[width=0.49\textwidth]{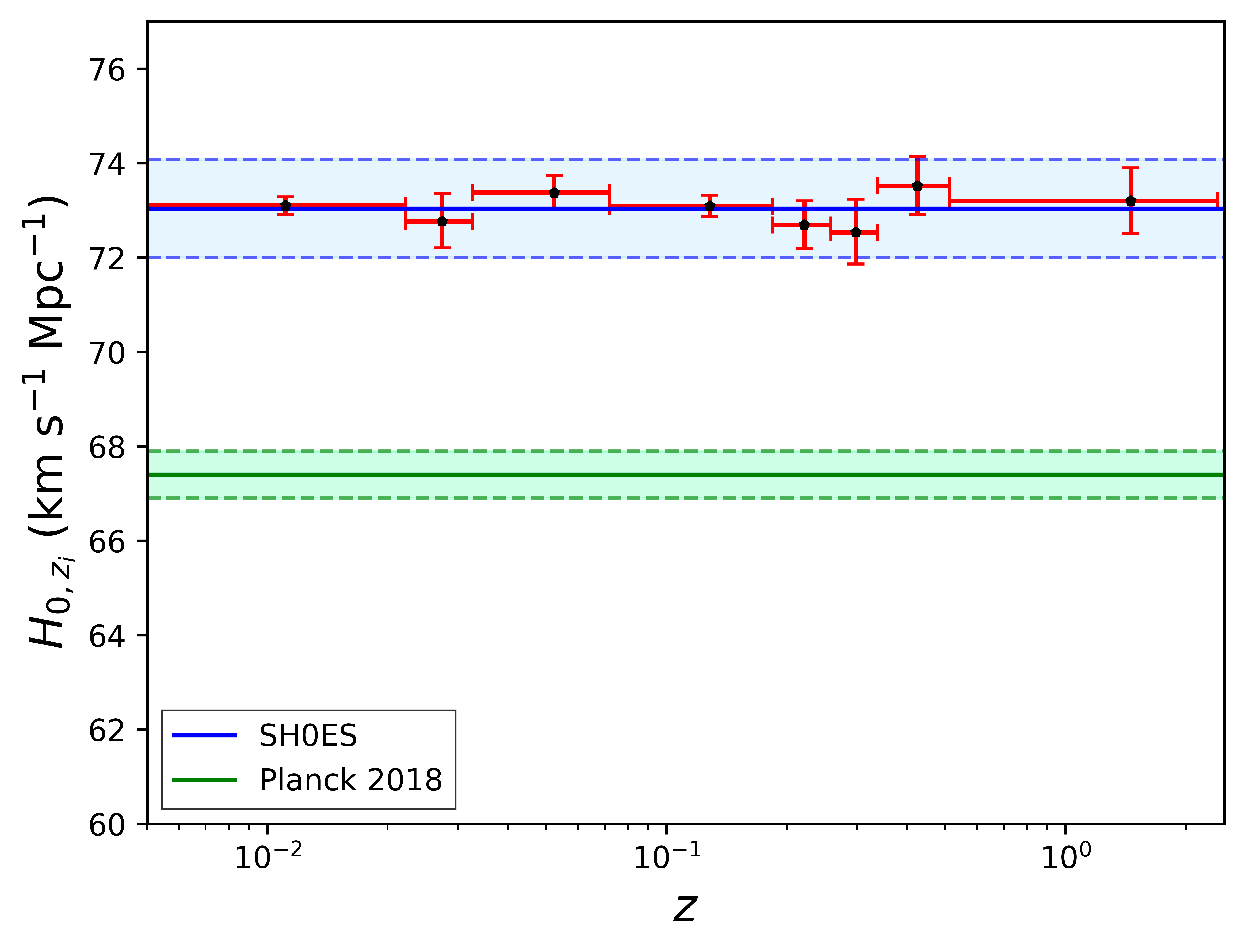}
\end{subfigure}
\begin{subfigure}
    \centering 
    \includegraphics[width=0.49\textwidth]{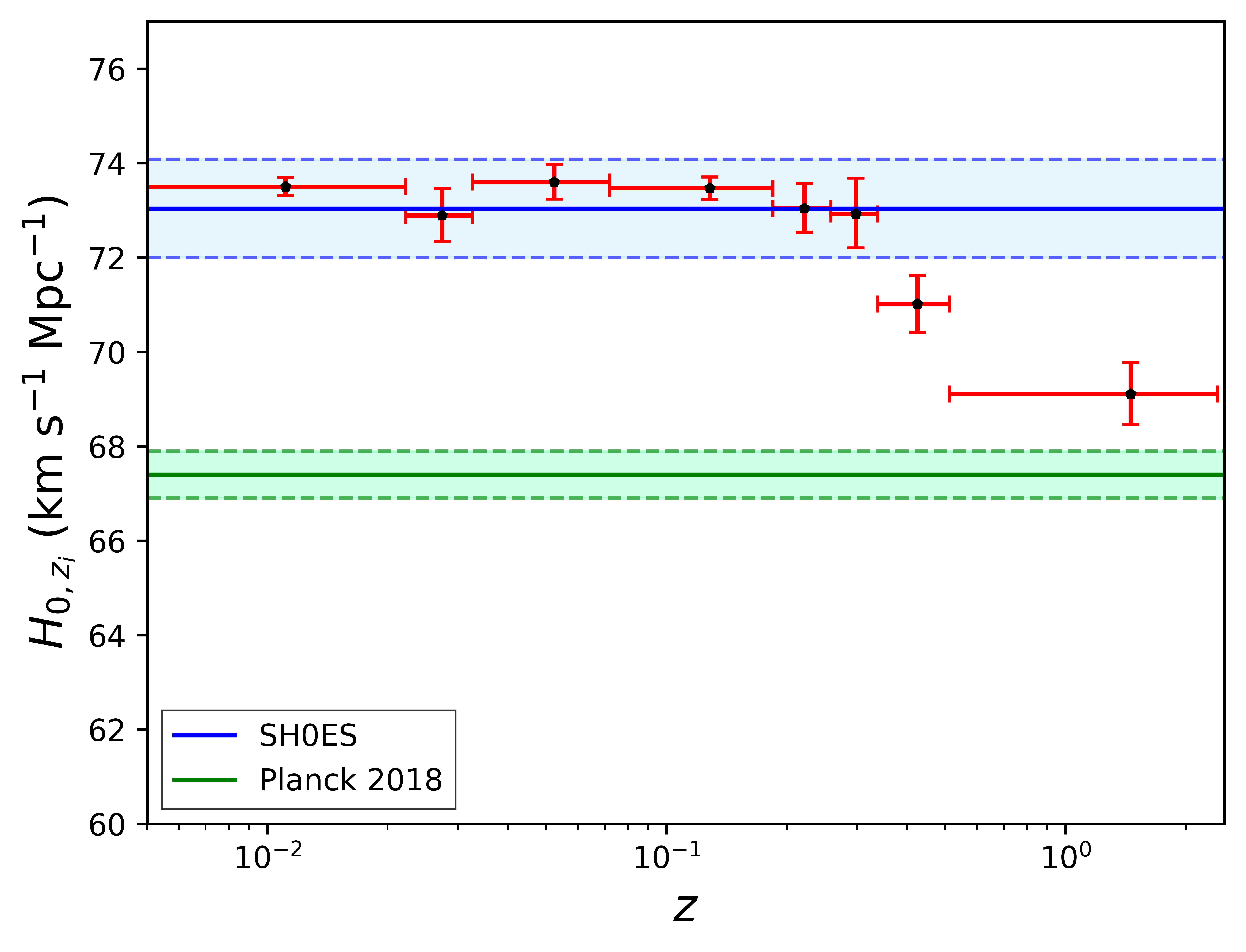}
\end{subfigure}
\caption{Uncorrelated $H_{0,z_i}$ posterior values with 68\% confidence level errors, for the \gdcdm\ (left) and the \lcdm\ (right) models, in the case of equal-number bins of CC+eBOSS+Pan+ dataset compared to the observational constraints on $H_0$ from SH0ES (blue) \cite{Riess:2021jrx} and Planck (green) \cite{Planck:2018vyg} collaborations.}
\label{benhp}
\end{figure}

In the equal-number binning method, we divide the data into eight bins, with, on average, 200 data points in each bin, as summarized in Table \ref{enbmb}. In Bayesian analysis, there is only one free parameter in each bin: $H_{0,z_i}$ for $i=1,\hdots,8$. Uniform prior distribution $50\le H_{0,z_i} \le 80$ ($km/s/Mpc$) is used in each bin. After obtaining the posterior median values from the Bayesian analysis, we removed the correlations of $H_{0,z_i}$ with one another by diagonalizing the covariance matrix as described in \cite{Jia:2022ycc}. We present the uncorrelated results for each bin in Table \ref{enbmb} for both the cases of the \gdcdm\ and the \lcdm\ models. We also plotted the uncorrelated posterior values of $H_{0,z_i}$ with 68\% confidence level errors in Fig. \ref{benhp}, together with the observational constraints on $H_0$ from the SH0ES (blue) \cite{Riess:2021jrx} and Planck (green) \cite{Planck:2018vyg} collaborations. We note that increasing the number of bins will result, on average, in a lower number of points in each bin. This will cause the uncertainty in $H_{0,z_i}$ to be more than $1.0$, which we try to avoid in the case of equal-number method.

\begin{table}[h]
\centering
\def\arraystretch{1.5}
\begin{tabular}{lccc}
\hline
\hline
Redshift bin & \vspace{1mm}\parbox[t]{15mm}{\centering Data number}
& \parbox[t]{18mm}{\centering \gdcdm\ $\mathbf{H_{0,z_i}}$}
& \parbox[t]{18mm}{\centering \lcdm\ $\mathbf{H_{0,z_i}}$} \\
\hline
$[0.0 - 0.0222]$ & $200$ & $73.10 ^{+0.19}_{-0.18}$ & $ 73.50^{+0.19}_{-0.19}$ \\
$[0.0222 - 0.0326]$ & $199$ & $72.77 ^{+0.59}_{-0.56}$ & $ 72.89 ^{+0.58}_{-0.55}$ \\
$[0.0326 - 0.072]$ & $200$ & $73.38 ^{+0.36}_{-0.36}$ & $ 73.60 ^{+0.37}_{-0.36}$ \\
$[0.072 - 0.1846]$ & $203$ & $73.09 ^{+0.23}_{-0.23}$ & $ 73.47 ^{+0.24}_{-0.24}$ \\
$[0.1846 - 0.258]$ & $201$ & $72.69 ^{+0.51}_{-0.49}$ & $ 73.04 ^{+0.53}_{-0.50}$ \\
$[0.258 - 0.338]$ & $201$ & $72.54 ^{+0.70}_{-0.67}$ & $ 72.92 ^{+0.76}_{-0.72}$ \\
$[0.338 - 0.512]$ & $213$ & $73.52 ^{+0.63}_{-0.61}$ & $ 71.02 ^{+0.61}_{-0.60}$ \\
$[0.512 - 2.400]$ & $217$ & $73.20 ^{+0.70}_{-0.69}$ & $ 69.11 ^{+0.67}_{-0.65}$ \\
\hline
\hline
\end{tabular}
\caption{Uncorrelated $H_{0,z_i}$ posterior values in each of the equal-number bins of the CC+eBOSS+Pan+ dataset for the \gdcdm\ and the \lcdm\ models.}
\label{enbmb}
\end{table}

\begin{figure}[h!]
\centering
    \includegraphics[width=0.80\textwidth]{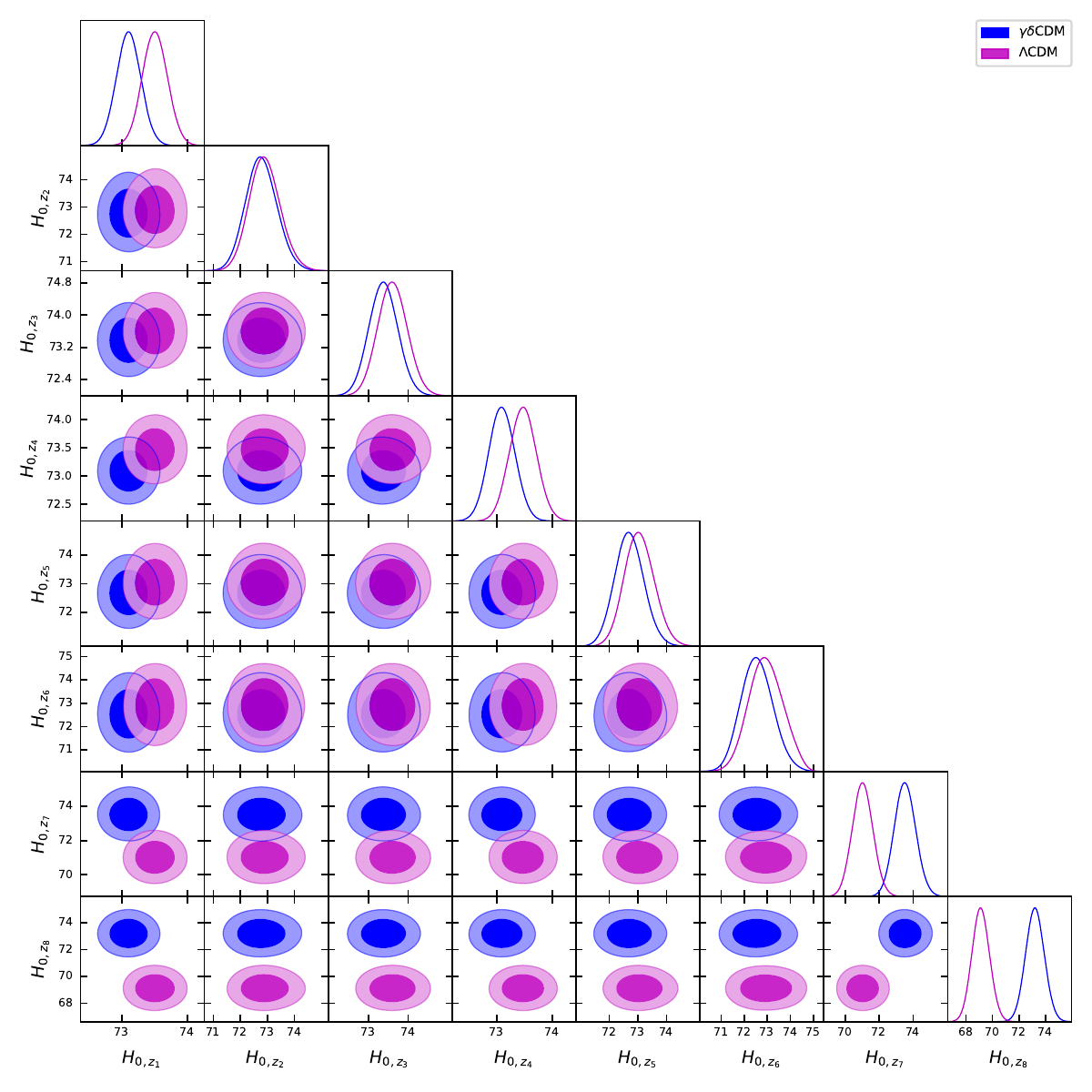}
\caption{Combined contour plots for 2D joint posterior distributions, with 1$\sigma$ and 2$\sigma$ confidence regions, for the $H_{0,z_i}$ in each equal-number bin of the CC+eBOSS+Pan+ dataset in the \gdcdm\ (blue) and the \lcdm\ (purple) models, together with 1D marginalized posterior distributions.}
\label{bencp}
\end{figure}

As seen in Table \ref{enbmb} and Fig. \ref{benhp}, there is a decreasing trend of the Hubble constant values obtained from the high redshift 7th and 8th bins in the case of \lcdm\ model, which is curiously missing in the \gdcdm\ model. The Hubble constant values obtained in each bin in the \gdcdm\ model are in the 68\% confidence interval of the late-time SH0ES observational result \cite{Riess:2021jrx}. The fact that the Hubble constant does not run with redshift up to redshift $z=2.4$ in the \gdcdm\ model indicates that $H_0$ is a true constant in the \gdcdm\ model. In contrast, the running of $H_0$ with redshift indicates that $H_0$ is not a true constant in the \lcdm\ model, which is paradoxical. Thus, this analysis points out the breakdown of the \lcdm\ model.

In Fig. \ref{bencp} we present combined contour plots for 2D joint posterior distributions, with 1$\sigma$ and 2$\sigma$ confidence regions, for $H_{0,z_i}$ in each equal-number bin of the CC+eBOSS+Pan+ dataset in the \gdcdm\ and the \lcdm\ models, together with 1D marginalized posterior distributions. This figure is also very illuminating in distinguishing the trends of the values of $H_{0,z_i}$ with respect to redshift in the \gdcdm\ and the \lcdm\ models. Both the 2D and the 1D posterior distributions visually show that the constant $H_0$ paradoxically runs with redshift in the \lcdm\ model, whereas it remains a constant in the \gdcdm\ model.

\subsubsection{Binning with Equal-Width Method \label{bewm}} 

\begin{figure}[h!]
\centering
\begin{subfigure}
    \centering 
    \includegraphics[width=0.49\textwidth]{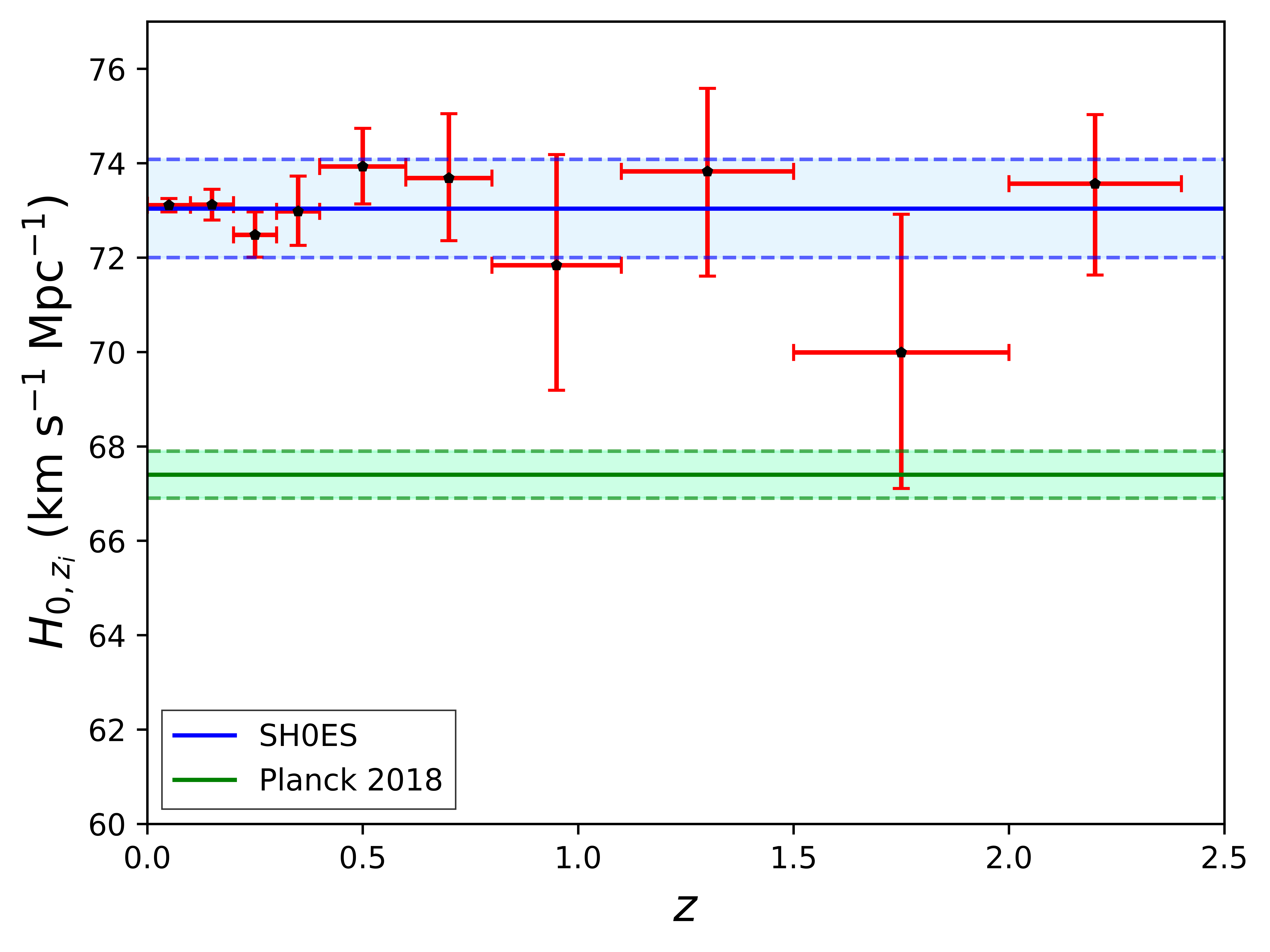}
\end{subfigure}
\begin{subfigure}
    \centering 
    \includegraphics[width=0.49\textwidth]{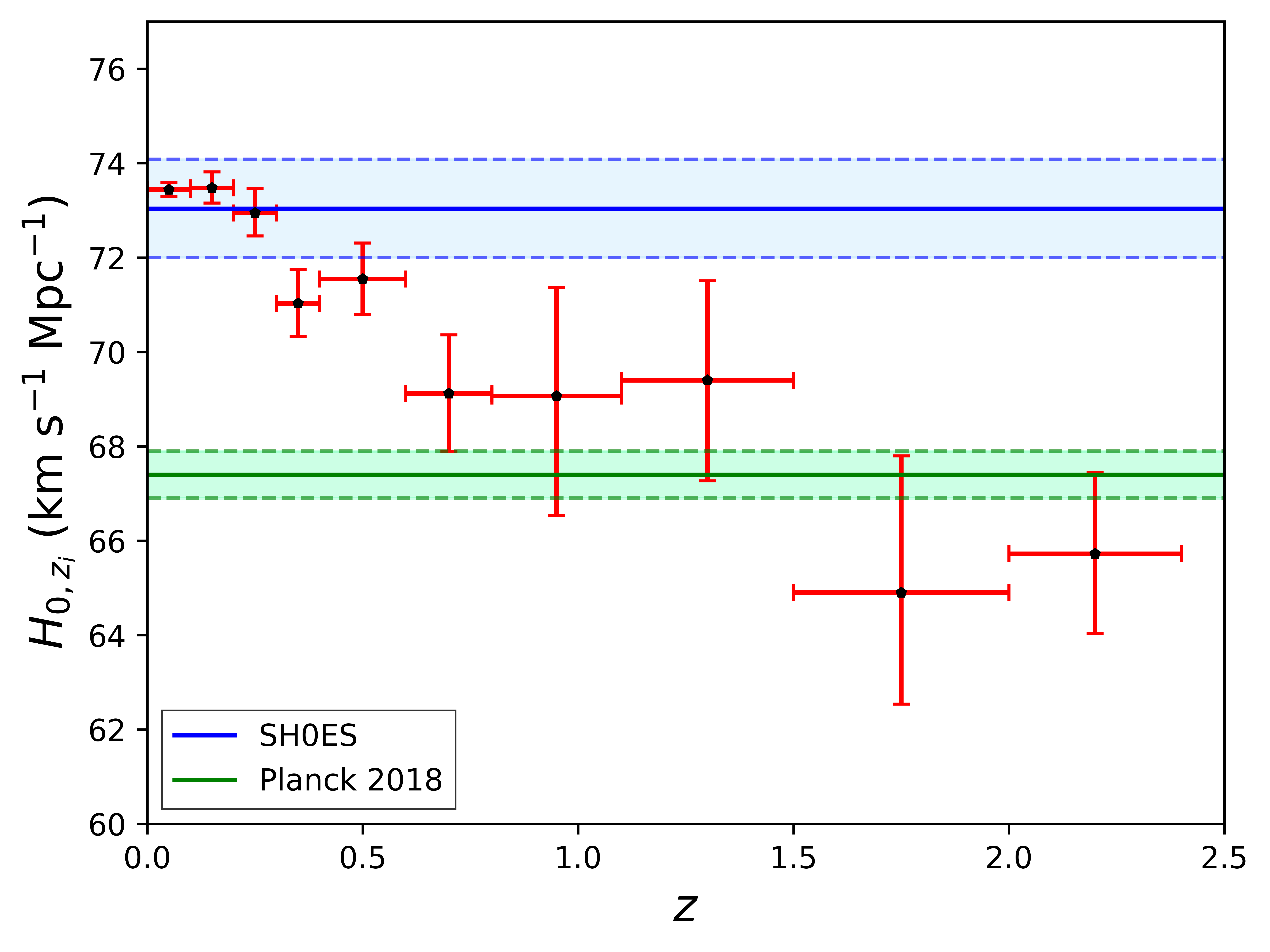}
\end{subfigure}
\caption{Uncorrelated $H_{0,z_i}$ posterior values with 68\% confidence level errors, for the \gdcdm\ (left) and the \lcdm\ (right) models, in the case of equal-width bins of CC+eBOSS+Pan+ dataset compared to the observational constraints on $H_0$ from SH0ES (blue) \cite{Riess:2021jrx} and Planck (green) \cite{Planck:2018vyg} collaborations.}
\label{bewhp}
\end{figure}

In the equal-width binning method, we divide the data into ten bins, the widths being $\Delta z = 0.1$ for the first four bins and $\Delta z = 0.4$ or $\Delta z = 0.5$ for the last three bins. Bins in the mid-$z$ range have widths $\Delta z = 0.2$ or $\Delta z = 0.3$. The width of the bins are increased with the redshift so that there are enough data points in each bin, so that the statistical analysis makes sense. Even then, there are too few data points in the last two bins. We chose the same bin sizes and the same number of bins to be able to compare our results with the results of \cite{Jia:2022ycc}. In the Bayesian analysis, there is only one free parameter in each bin: $H_{0,z_i}$ for $i=1,\hdots,10$. Uniform prior distribution $50\le H_{0,z_i} \le 80$ ($km/s/Mpc$) is used in each bin. After obtaining the posterior median values from the Bayesian analysis, we removed the correlations of $H_{0,z_i}$ with one another by diagonalizing the covariance matrix as described in \cite{Jia:2022ycc}. We present the uncorrelated results for each bin in Table \ref{ewbmb} for both the cases of the \gdcdm\ and the \lcdm\ models. We also plotted the uncorrelated posterior values of $H_{0,z_i}$ with 68\% confidence level errors in Fig. \ref{bewhp}, together with the observational constraints on $H_0$ from the SH0ES (blue) \cite{Riess:2021jrx} and Planck (green) \cite{Planck:2018vyg} collaborations.

\begin{table}[h]
\centering
\def\arraystretch{1.5}
\begin{tabular}{lccc}
\hline
\hline
Redshift bin & \vspace{1mm}\parbox[t]{15mm}{\centering Data number}
& \parbox[t]{18mm}{\centering \gdcdm\ $\mathbf{H_{0,z_i}}$}
& \parbox[t]{18mm}{\centering \lcdm\ $\mathbf{H_{0,z_i}}$} \\
\hline
$[0.0 - 0.1]$ & $632$ & $73.11 ^{+0.14}_{-0.14}$ & $ 73.44 ^{+0.14}_{-0.14}$ \\
$[0.1 - 0.2]$ & $212$ & $73.12 ^{+0.33}_{-0.32}$ & $ 73.48 ^{+0.34}_{-0.32}$ \\
$[0.2 - 0.3]$ & $262$ & $72.48 ^{+0.49}_{-0.47}$ & $ 72.95 ^{+0.51}_{-0.49}$ \\
$[0.3 - 0.4]$ & $190$ & $72.98 ^{+0.75}_{-0.72}$ & $ 71.03 ^{+0.72}_{-0.70}$ \\
$[0.4 - 0.6]$ & $189$ & $73.93 ^{+0.81}_{-0.79}$ & $ 71.55 ^{+0.76}_{-0.75}$ \\
$[0.6 - 0.8]$ & $104$ & $73.69 ^{+1.36}_{-1.33}$ & $ 69.12 ^{+1.24}_{-1.22}$ \\
$[0.8 - 1.1]$ & $15$ & $71.84 ^{+2.35}_{-2.65}$ & $ 69.07 ^{+2.30}_{-2.53}$ \\
$[1.1 - 1.5]$ & $18$ & $73.83 ^{+1.76}_{-2.22}$ & $ 69.40 ^{+2.11}_{-2.13}$ \\
$[1.5 - 2.0]$ & $9$ & $70.00 ^{+2.93}_{-2.88}$ & $ 64.90 ^{+2.90}_{-2.36}$ \\
$[2.0 - 2.4]$ & $3$ & $73.57 ^{+1.47}_{-1.94}$ & $ 65.72 ^{+1.73}_{-1.69}$ \\
\hline
\hline
\end{tabular}
\caption{$H_{0,z_i}$ posterior values in each of the equal-width bins of the CC+eBOSS+Pan+ dataset for the \gdcdm\ and the \lcdm\ models.}
\label{ewbmb}
\end{table}

\begin{figure}[h!]
\centering
    \includegraphics[width=0.80\textwidth]{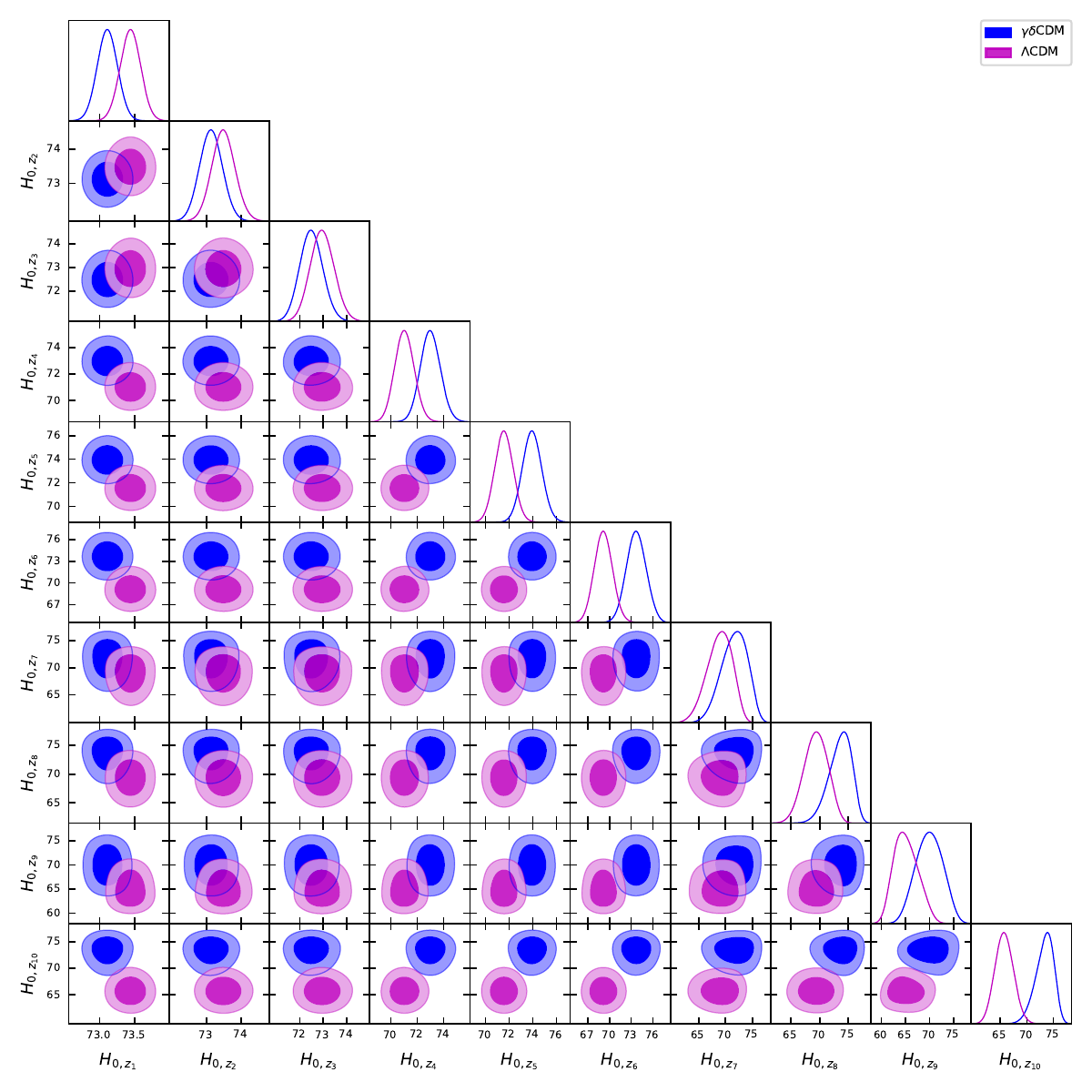}
\caption{Combined contour plots for 2D joint posterior distributions, with 1$\sigma$ and 2$\sigma$ confidence regions, for the $H_{0,z_i}$ in each equal-width bin of the CC+BAO+Pan+ dataset in the \gdcdm\ (blue) and the \lcdm\ (purple) models, together with 1D marginalized posterior distributions.}
\label{bewcp}
\end{figure}

As seen in Table \ref{ewbmb} and Fig. \ref{bewhp}, there is a ``clear'' decreasing trend of the Hubble constant values in the case of \lcdm\ model, which is curiously missing again in the \gdcdm\ model. The last two bins contain too few data points and therefore the results of the Bayesian analysis are not as trustable as in the case of other bins. The low value obtained in the 9th bin in both models might indicate a systematic problem with some data in that bin. The Hubble constant values obtained in each bin in the \gdcdm\ model are in the 68\% confidence interval, or very close to the late time SH0ES observational result \cite{Riess:2021jrx}. The fact that the Hubble constant does not have an obvious change with redshift up to redshift $z=2.4$ in the \gdcdm\ model indicates that $H_0$ is a true constant in the \gdcdm\ model. In contrast, the running of $H_0$ with redshift indicates that $H_0$ is not a true constant in the \lcdm\ model, which is paradoxical. Thus, this analysis also points out the breakdown of the \lcdm\ model.

In Fig. \ref{bewcp} we present combined contour plots for 2D joint posterior distributions, with 1$\sigma$ and 2$\sigma$ confidence regions, for $H_{0,z_i}$ in each equal-width bin of the CC+eBOSS+Pan+ dataset in the \gdcdm\ and the \lcdm\ models, together with 1D marginalized posterior distributions. This figure is also very illuminating to distinguish the trends of $H_{0,z_i}$ values with respect to redshift in the \gdcdm\ and the \lcdm\ models. Both the 2D and the 1D posterior distributions visually show that the constant $H_0$ paradoxically runs with redshift in the \lcdm\ model, whereas it remains a constant in the \gdcdm\ model.

\subsubsection{Comparison of the cosmological models \label{comb}} 

It is expected that a model with extra free parameters would have a higher maximum value of the likelihood function (\ref{like}) compared to the model with a lower number of parameters. The model comparison criteria takes into account this disparity by penalizing the existence of extra free parameters. The most widely used information criteria are the Akaike information criterion (AIC) \cite{Akaike:1974vps} and the Bayesian information criterion (BIC) \cite{Schwarz:1978vps}. The AIC is defined by AIC $= 2k - 2\ln\mathcal{L}$ and the BIC is defined by BIC $= k\ln N - 2\ln\mathcal{L}$, where $k$ is the number of model parameters, $N$ is the number of data points, and $\mathcal{L}$ is the maximum value of the likelihood function (\ref{like}). 
For large dataset sizes, the penalty is larger in BIC than in AIC \cite{Stoica:2004vps}.

The \lcdm\ model has only one free parameter, $H_0$, after the matter density parameter is fixed (see section \ref{fcbp}). The \gdcdm\ model has two extra free cosmological parameters compared to the \lcdm\ model. In the analysis for the \gdcdm\ model with the full CC+eBOSS+Pan+ dataset we did not fix the matter density parameter, and therefore this analysis is done with four free parameters.
In the analysis with equal-number bins there are eight bins, and for each bin there is one free parameter $H_{0,z_i}$. In the analysis with equal-width bins, there are ten bins and for each bin there is one free parameter $H_{0,z_i}$. 

In Table \ref{icb} we present $min\ \chi^2$, AIC and BIC values calculated for all the analyses done with distinct binning methods and with distinct models. We also present differences in the information criteria values, calculated in each of the full or distinctly binned datasets separately. $\Delta \mathrm{AIC}$ is defined by $\Delta \mathrm{AIC} = \mathrm{AIC}_{model} - \mathrm{AIC}_{\ga\de CDM}$ for each distinct model and similarly for $\Delta$BIC.

The differences in the values of the information criteria in each of the datasets are noteworthy, as presented in Table \ref{icb}. Although \gdcdm\ is penalized for having additional parameters, the information criteria strongly favor it over the \lcdm\ model.  The \gdcdm\ model with constant $H_0$ is also favored over the binned models. This is in contrast with the case presented in \cite{Jia:2022ycc} for the \lcdm\ model. Binned analysis with redshift dependent $H_0$ is favored over the \lcdm\ model with constant $H_0$ (see Table 5 in \cite{Jia:2022ycc}). Thus, we can safely conclude that the CC+eBOSS+Pan+ late-time observational dataset favors the \gdcdm\ cosmological model for which $H_0$ is a true constant.

\begin{table}[hbt!]
\centering
\def\arraystretch{1.5}
\begin{tabular}{l|lccccc}
\hline
\hline
\parbox[c]{20mm}{Dataset} & \parbox[c]{12mm}{Model} 
& \parbox[c]{12mm}{\centering $\mathbf{min\ \chi^2}$} 
& \parbox[c]{12mm}{\centering $\mathbf{AIC}$} 
& \parbox[c]{12mm}{\centering $\mathbf{\Delta AIC}$} 
& \parbox[c]{12mm}{\centering $\mathbf{BIC}$} 
& \parbox[c]{12mm}{\centering $\mathbf{\Delta BIC}$} \\
\hline
Full dataset & \gdcdm\ & $1429.0$  & $1437.0$ & 0 & $1458.6$ & 0 \\
& \lcdm\ & $1568.2$  & $1570.2$ & 133.2 & $1575.6$ & 117 \\
\hline
Equal-number bins & \gdcdm\ & $1427.4$  & $1443.4$ & 6.4 & $1486.6$ & 28.0 \\
& \lcdm\ & $1520.6$  & $1536.6$ & 99.6 & $1579.8$ & 121.2 \\
\hline
Equal-width bins & \gdcdm\ & $1426.2$  & $1446.2$ & 9.2 & $1500.2$ & 41.6 \\
& \lcdm\ & $1514.5$  & $1534.5$ & 97.5 & $1588.5$ & 129.9 \\
\hline
\hline
\end{tabular}
\caption{Minimum chi-squared, Akaike Information Criterion (AIC) and Bayesian Information Criterion (BIC) results for the \gdcdm\ and the \lcdm\ models fitted to the CC+eBOSS+Pan+ dataset.  Differences of information criterion values are calculated with $\Delta \mathrm{AIC} = \mathrm{AIC}_{model} - \mathrm{AIC}_{\ga\de CDM}$ for each distinct model and similarly for $\Delta$BIC.}
\label{icb}
\end{table}

\subsection{Analysis with the CC+DESI+Pantheon+ Dataset \label{cbp}} 

The CC+DESI+Pantheon+ dataset include 32 cosmic chronometers Hubble data points, 12 new BAO measurement data points obtained by DESI collaboration, and  1590 light curves of distinct Type Ia supernovae. The details of individual datasets are given in Section \ref{data}. We analyzed both the \lcdm\ and \gdcdm\ models firstly with the full dataset, then with the binned dataset of 8 bins with equal number data points and finally with the binned dataset of 10 bins with equal redshift width.

\subsubsection{Analysis with the full dataset \label{fcbdp}} 

When we constrain the flat \lcdm\ model with the full CC+DESI+Pan+ dataset, with the choice $\Omega_{m0}=0.3$ and uniform prior distribution $50\le H_{0,z_i} \le 80$ ($km/s/Mpc$) for all $z_i$, the Bayesian analysis resulted in a posterior value for the only free parameter to be $H_0 = 73.06\pm 0.12\ km/s/Mpc$. In Bayesian analysis with the BAO data, the sound horizon at the baryon drag is taken to be $r_d =  147.1$ Mpc \cite{Planck:2018vyg} for the \lcdm\ model. 

\begin{figure}[h!]
\centering
   \includegraphics[width=0.46\textwidth]{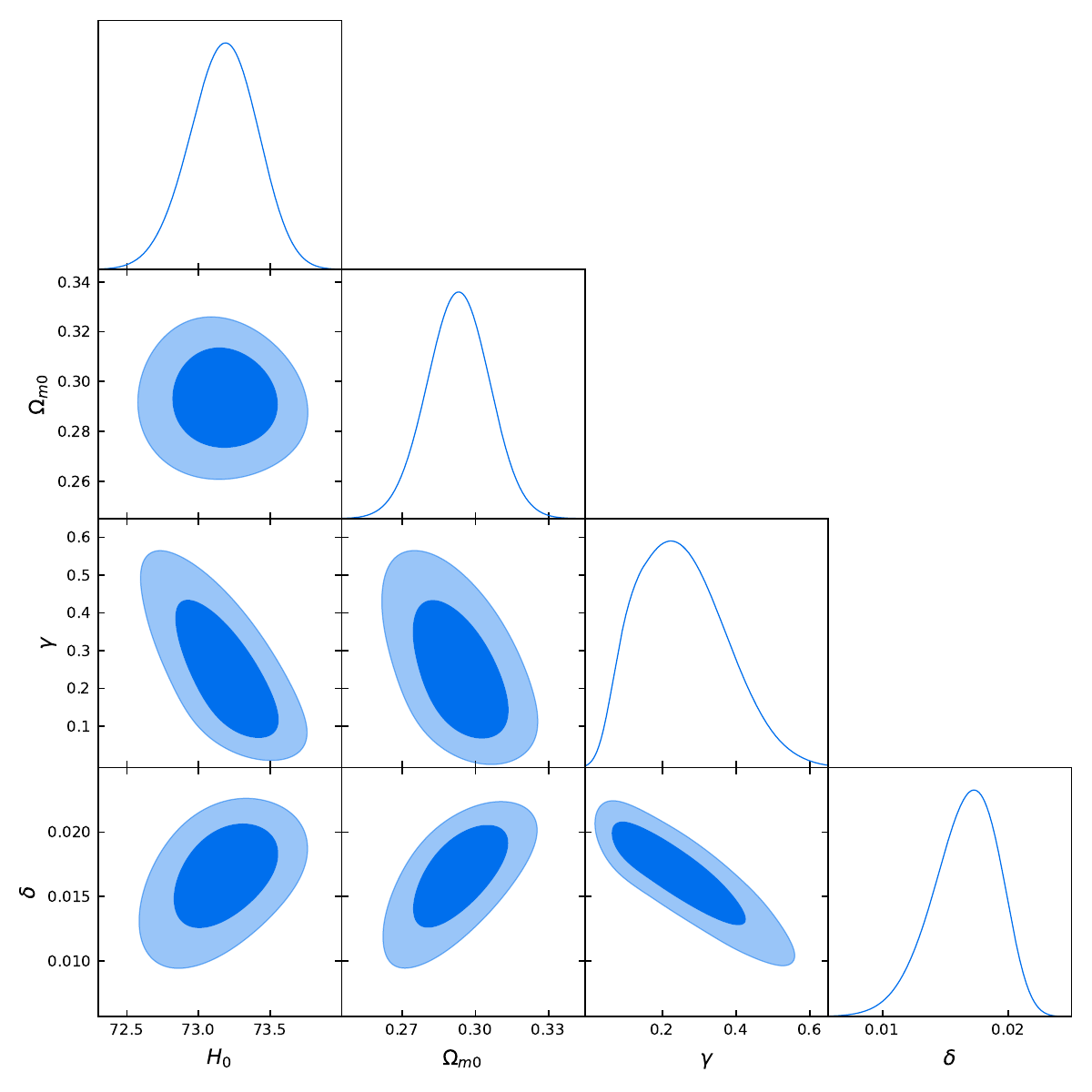}
\caption{The contour plots for 2D joint posterior distributions, with 1$\sigma$ and 2$\sigma$ confidence regions, for the free cosmological parameters ($H_0,\ \Omega_{m0},\ \gamma$ and $\delta$) of the \gdcdm\ model for the full CC+DESI+Pan+ dataset, together with 1D marginalized posterior distributions.}
\label{fullgdd}
\end{figure}

Constraining the \gdcdm\ model with the full CC+DESI+Pan+ dataset is done with the same choices as in Section \ref{fcbp}.
Bayesian analysis resulted in the following median posterior values with $1\sigma$ confidence intervals for the cosmological parameters: $H_0 = 73.18\pm 0.24\ km/s/Mpc$, $\Omega_{m0} = 0.293\pm 0.013$, $\ga = 0.242\pm 0.13$ and $\de = 0.017\pm 0.003$. In Fig. \ref{fullgdd} we present the marginalized 1D and 2D posterior distributions of the cosmological parameters of the \gdcdm\ model.
In the Bayesian analyses with the binned datasets we fixed $\Omega_{m0}$, $\ga$ and $\de$ to these values and left only the parameters $H_{0,z_i}$ in each bin free with uniform prior distribution $50\le H_{0,z_i} \le 80$ ($km/s/Mpc$) for all $z_i$. The sound horizon in the case of the \gdcdm\ model is calculated from the values of the cosmological parameters from the full dataset fit and is found to be  $r_d = 137.1$ Mpc. This value of the sound horizon is also in the 1$\sigma$ range of the model-independent estimate from a recalibration of eBOSS and DESI BAO datasets, obtained by deep learning techniques \cite{Shah:2024gfu}.

\subsubsection{Binning with Equal-Number Method \label{denm}} 

\begin{figure}[h!]
\centering
\begin{subfigure}
    \centering 
    \includegraphics[width=0.49\textwidth]{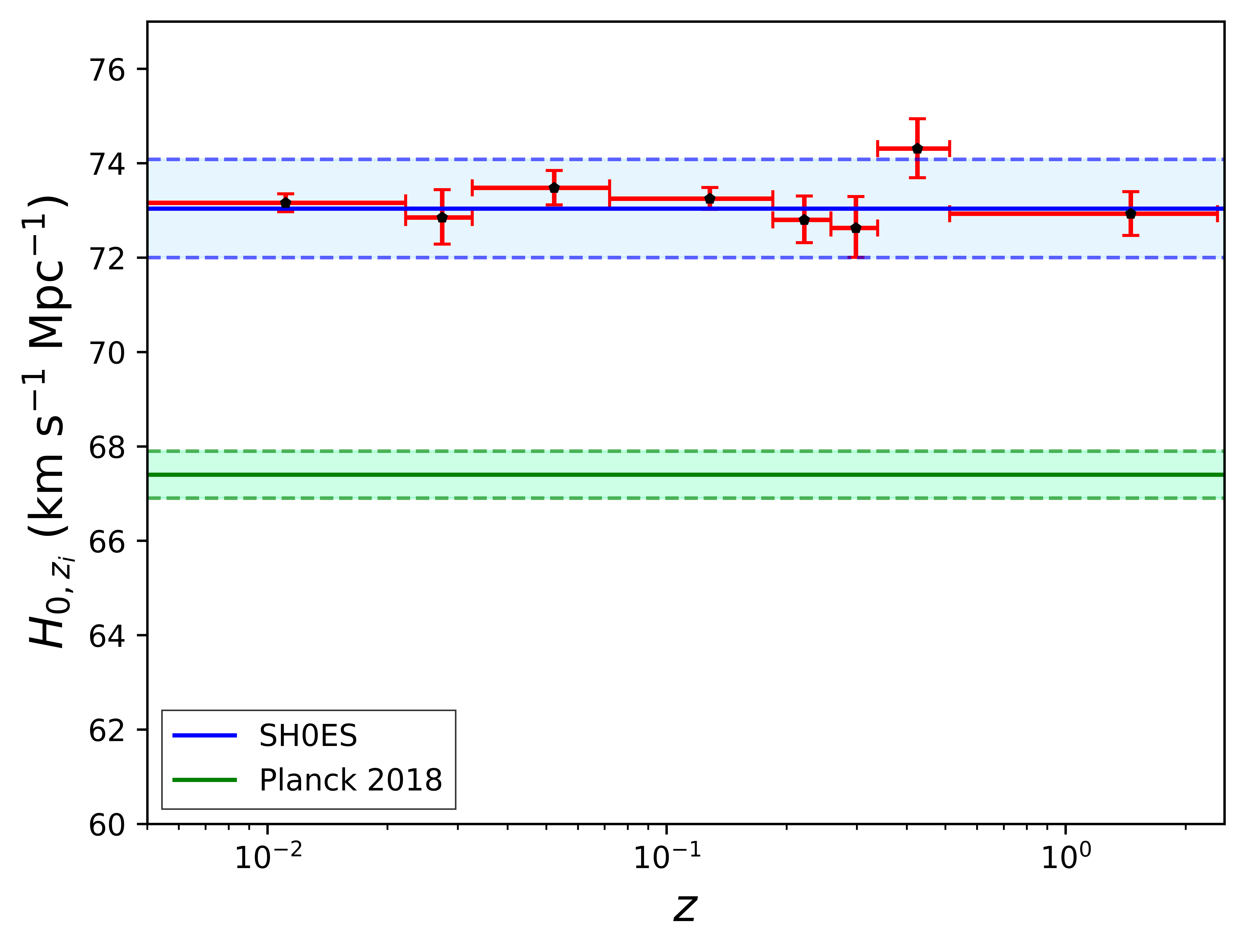}
\end{subfigure}
\begin{subfigure}
    \centering 
    \includegraphics[width=0.49\textwidth]{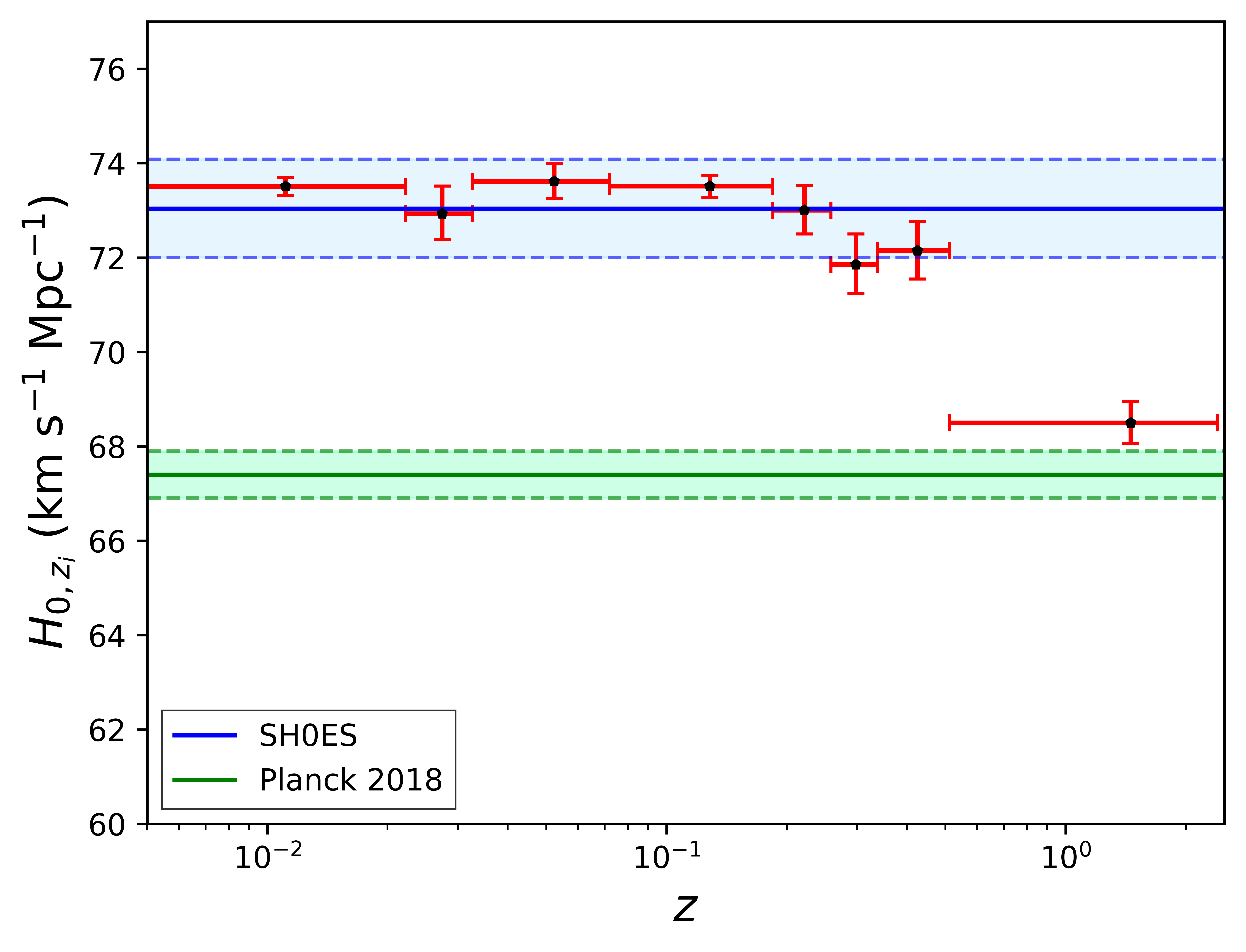}
\end{subfigure}
\caption{Uncorrelated $H_{0,z_i}$ posterior values with 68\% confidence level errors, for the \gdcdm\ (left) and the \lcdm\ (right) models, in the case of equal-number bins of CC+DESI+Pan+ dataset compared to the observational constraints on $H_0$ from SH0ES (blue) \cite{Riess:2021jrx} and Planck (green) \cite{Planck:2018vyg} collaborations.}
\label{denhp}
\end{figure}

We divide the data into bins with equal number of data points, and perform data analysis as described in section \ref{benm}. We present the uncorrelated results for each bin in Table \ref{enbmd} for both the cases of the \gdcdm\ and the \lcdm\ models. We also plotted the uncorrelated posterior values of $H_{0,z_i}$ with 68\% confidence level errors in Fig. \ref{denhp}, together with the observational constraints on $H_0$ from the SH0ES (blue) \cite{Riess:2021jrx} and Planck (green) \cite{Planck:2018vyg} collaborations.

\begin{figure}[h!]
\centering
    \includegraphics[width=0.80\textwidth]{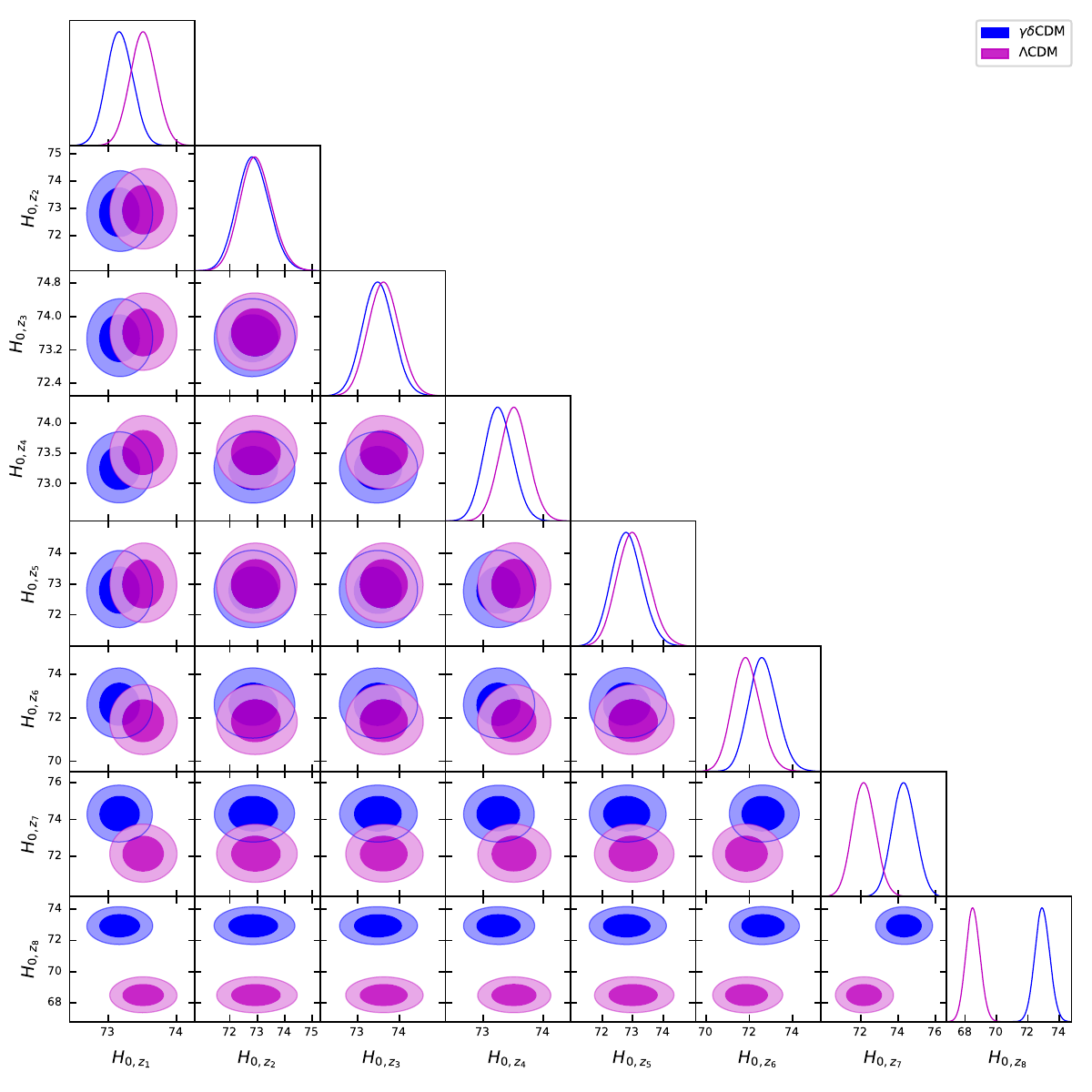}
\caption{Combined contour plots for 2D joint posterior distributions, with 1$\sigma$ and 2$\sigma$ confidence regions, for the $H_{0,z_i}$ in each equal-number bin of the CC+DESI+Pan+ dataset in the \gdcdm\ (blue) and the \lcdm\ (purple) models, together with 1D marginalized posterior distributions.}
\label{dencp}
\end{figure}

\begin{table}[h]
\centering
\def\arraystretch{1.5}
\begin{tabular}{lccc}
\hline
\hline
Redshift bin & \vspace{1mm}\parbox[t]{15mm}{\centering Data number}
& \parbox[t]{18mm}{\centering \gdcdm\ $\mathbf{H_{0,z_i}}$}
& \parbox[t]{18mm}{\centering \lcdm\ $\mathbf{H_{0,z_i}}$} \\
\hline
$[0.0 - 0.0222]$ & $200$ & $73.16 ^{+0.19}_{-0.19}$ & $ 73.51 ^{+0.19}_{-0.19}$ \\
$[0.0222 - 0.0326]$ & $199$ & $72.84 ^{+0.59}_{-0.56}$ & $ 72.93 ^{+0.58}_{-0.55}$ \\
$[0.0326 - 0.072]$ & $200$ & $73.48 ^{+0.37}_{-0.36}$ & $ 73.62 ^{+0.37}_{-0.36}$ \\
$[0.072 - 0.1846]$ & $202$ & $73.25 ^{+0.24}_{-0.23}$ & $ 73.51 ^{+0.24}_{-0.23}$ \\
$[0.1846 - 0.258]$ & $201$ & $72.80 ^{+0.51}_{-0.48}$ & $ 73.00 ^{+0.53}_{-0.50}$ \\
$[0.258 - 0.338]$ & $202$ & $72.61 ^{+0.67}_{-0.62}$ & $ 71.85 ^{+0.65}_{-0.61}$ \\
$[0.338 - 0.512]$ & $211$ & $74.31 ^{+0.64}_{-0.62}$ & $ 72.15 ^{+0.62}_{-0.60}$ \\
$[0.512 - 2.400]$ & $219$ & $72.93 ^{+0.48}_{-0.46}$ & $ 68.50 ^{+0.45}_{-0.44}$ \\
\hline
\hline
\end{tabular}
\caption{$H_{0,z_i}$ posterior values in each of the equal-number bins of the CC+DESI+Pan+ dataset for the \gdcdm\ and the \lcdm\ models.}
\label{enbmd}
\end{table}

As seen in Table \ref{enbmd} and Fig. \ref{denhp}, decreasing trend of the Hubble constant happens in the 8th bin in the case of \lcdm\ model, later than the case shown in Fig. \ref{benhp}. The new DESI BAO data have more relevance in the 7th bin and make the decreasing trend less pronounced compared to the dataset without DESI BAO data points.
We again do not observe any decreasing trend of $H_0$ in the \gdcdm\ model. The Hubble constant values obtained in each bin in the \gdcdm\ model are in the 68\% confidence interval of the late time SH0ES observational result \cite{Riess:2021jrx}. The fact that the Hubble constant does not run with redshift up to redshift $z=2.4$ in the \gdcdm\ model indicates, as before, that $H_0$ is a true constant in the \gdcdm\ model. In contrast, the running of $H_0$ with redshift indicates the inconsistency that $H_0$ is not a constant in the \lcdm\ model. This analysis also points out the breakdown of the \lcdm\ model as in Section \ref{benm}.

In Fig. \ref{dencp} we present combined contour plots for 2D joint posterior distributions, with 1$\sigma$ and 2$\sigma$ confidence regions, for $H_{0,z_i}$ in each equal-number bin of the CC+DESI+Pan+ dataset in the \gdcdm\ and the \lcdm\ models, together with 1D marginalized posterior distributions. This figure is also very illuminating to distinguish the trends of $H_{0,z_i}$ values with respect to redshift in the \gdcdm\ and the \lcdm\ models. Both the 2D and the 1D posterior distributions visually show that the constant $H_0$ paradoxically runs with redshift in the \lcdm\ model, whereas it remains a constant in the \gdcdm\ model.

\subsubsection{Binning with Equal-Width Method \label{dewm}} 

\begin{figure}[h!]
\centering
\begin{subfigure}
    \centering 
    \includegraphics[width=0.49\textwidth]{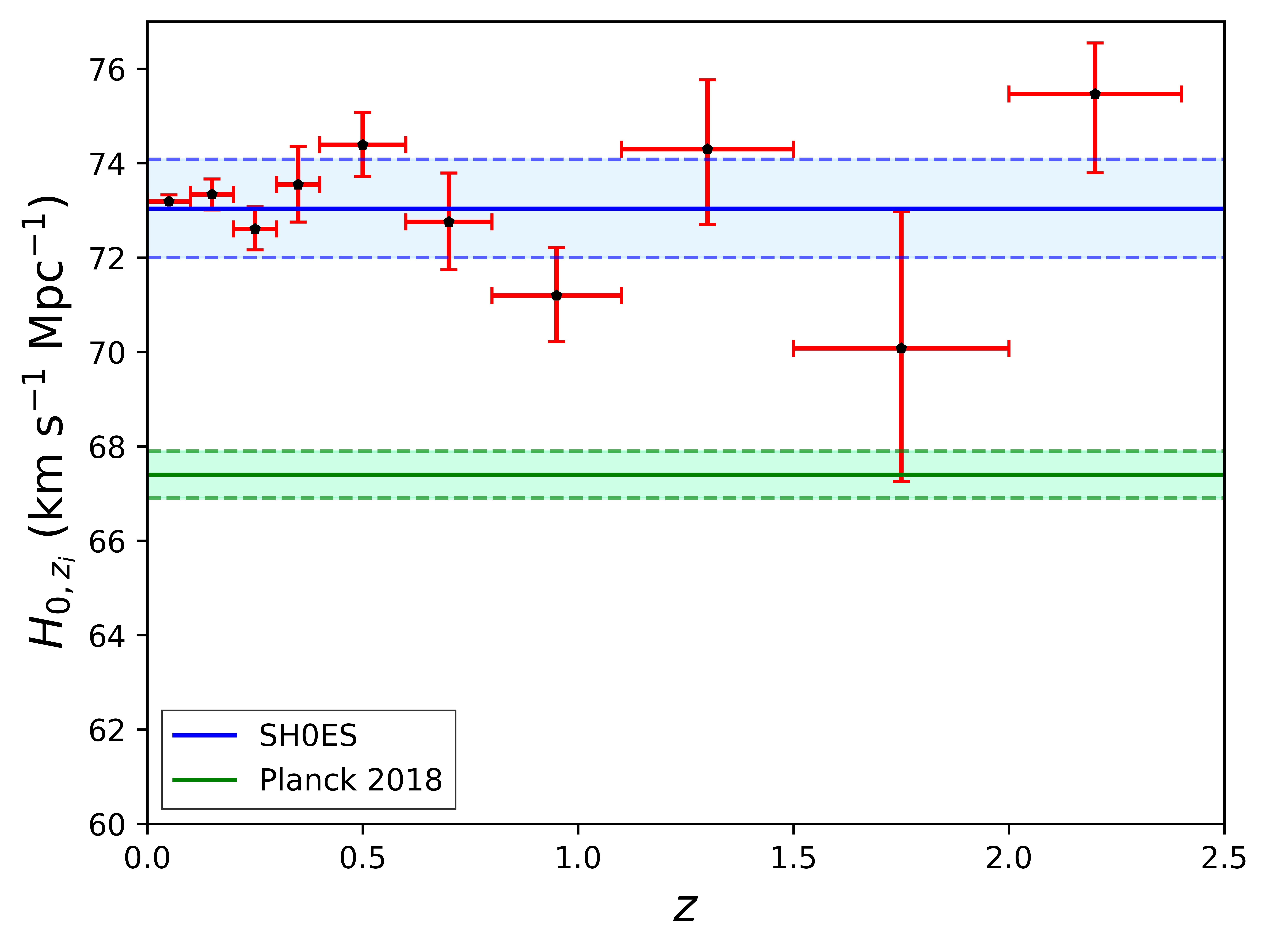}
\end{subfigure}
\begin{subfigure}
    \centering 
    \includegraphics[width=0.49\textwidth]{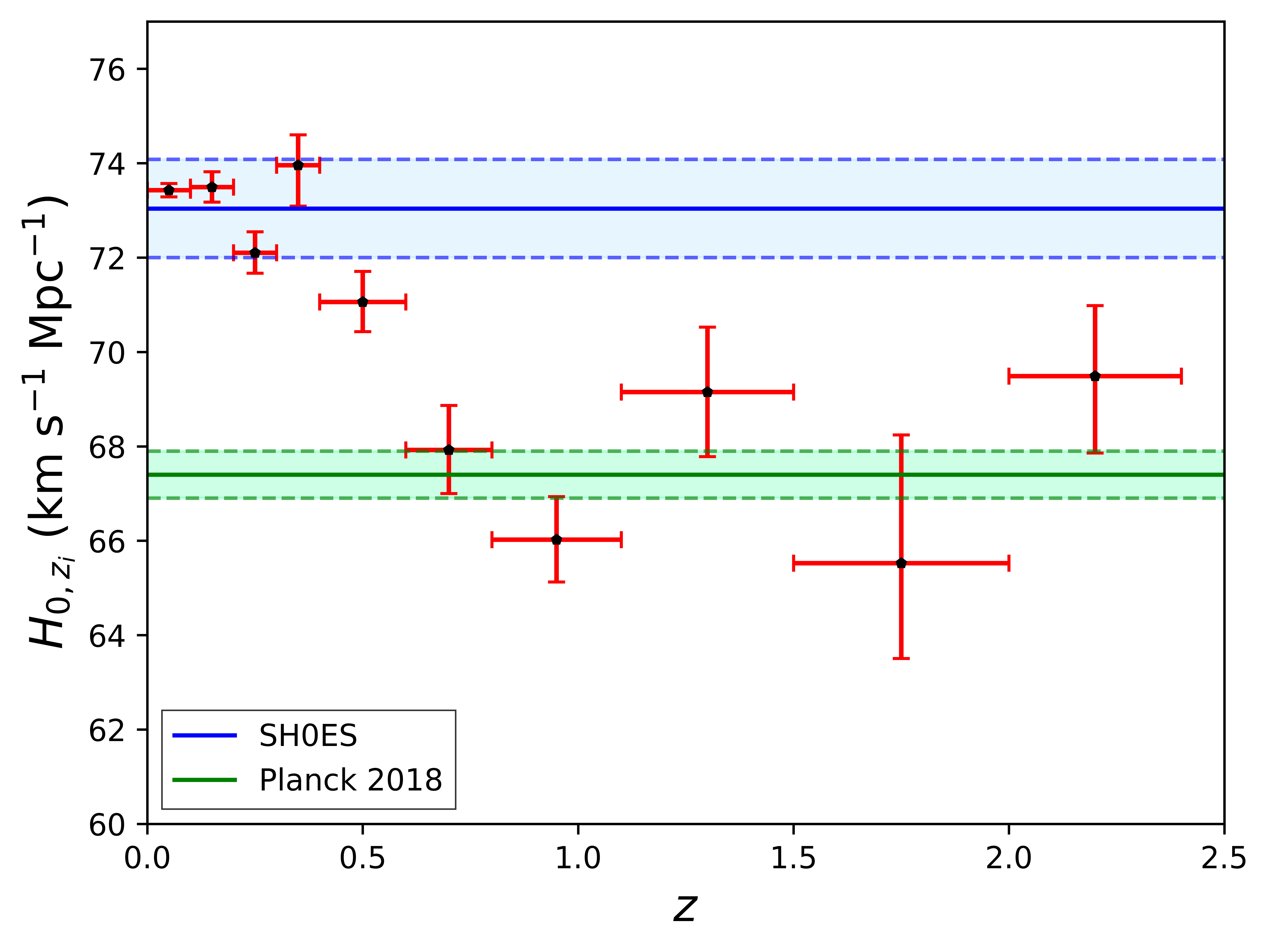}
\end{subfigure}
\caption{Uncorrelated $H_{0,z_i}$ posterior values with 68\% confidence level errors, for the \gdcdm\ (left) and the \lcdm\ (right) models, in the case of equal-width bins of CC+DESI+Pan+ dataset compared to the observational constraints on $H_0$ from SH0ES (blue) \cite{Riess:2021jrx} and Planck (green) \cite{Planck:2018vyg} collaborations.}
\label{dewhp}
\end{figure}

In the equal-width binning method we divide the data into ten bins, and perform the data analysis as described in Section \ref{bewm}. We present the uncorrelated results for each bin in Table \ref{ewbmd} for both the cases of the \gdcdm\ and the \lcdm\ models. We also plotted the uncorrelated posterior values of $H_{0,z_i}$ with 68\% confidence level errors in Fig. \ref{dewhp}, together with the observational constraints on $H_0$ from the SH0ES (blue) \cite{Riess:2021jrx} and Planck (green) \cite{Planck:2018vyg} collaborations.

As seen in Table \ref{ewbmd} and Fig. \ref{dewhp} decreasing trend of the Hubble constant values in the case of \lcdm\ model is much more clear in this dataset. Again in the \gdcdm\ model the decreasing trend is not observed. The last two bins contain too few data points and therefore the results of the Bayesian analysis are not as trustable as in the case of other bins. The two lowest values obtained in the 7th and the 9th bins in both models might indicate a systematic problem with some data in those bins. The Hubble constant values obtained in each bin in the \gdcdm\ model are in the 68\% confidence interval, or very close to the late time SH0ES observational result \cite{Riess:2021jrx}. The fact that the Hubble constant does not have an obvious change with redshift up to redshift $z=2.4$ in the \gdcdm\ model indicates that $H_0$ is a true constant in the \gdcdm\ model. In contrast, the running of $H_0$ with redshift contradicts the fact that $H_0$ should be a constant in the \lcdm\ model. Thus, this analysis also points out the breakdown of the \lcdm\ model. Our result for the \lcdm\ model fitted to the CC+DESI+Pan+ dataset in equal-width bins agrees with the result of the analysis performed in \cite{Jia:2024wix}.

\begin{figure}[h!]
\centering
    \includegraphics[width=0.80\textwidth]{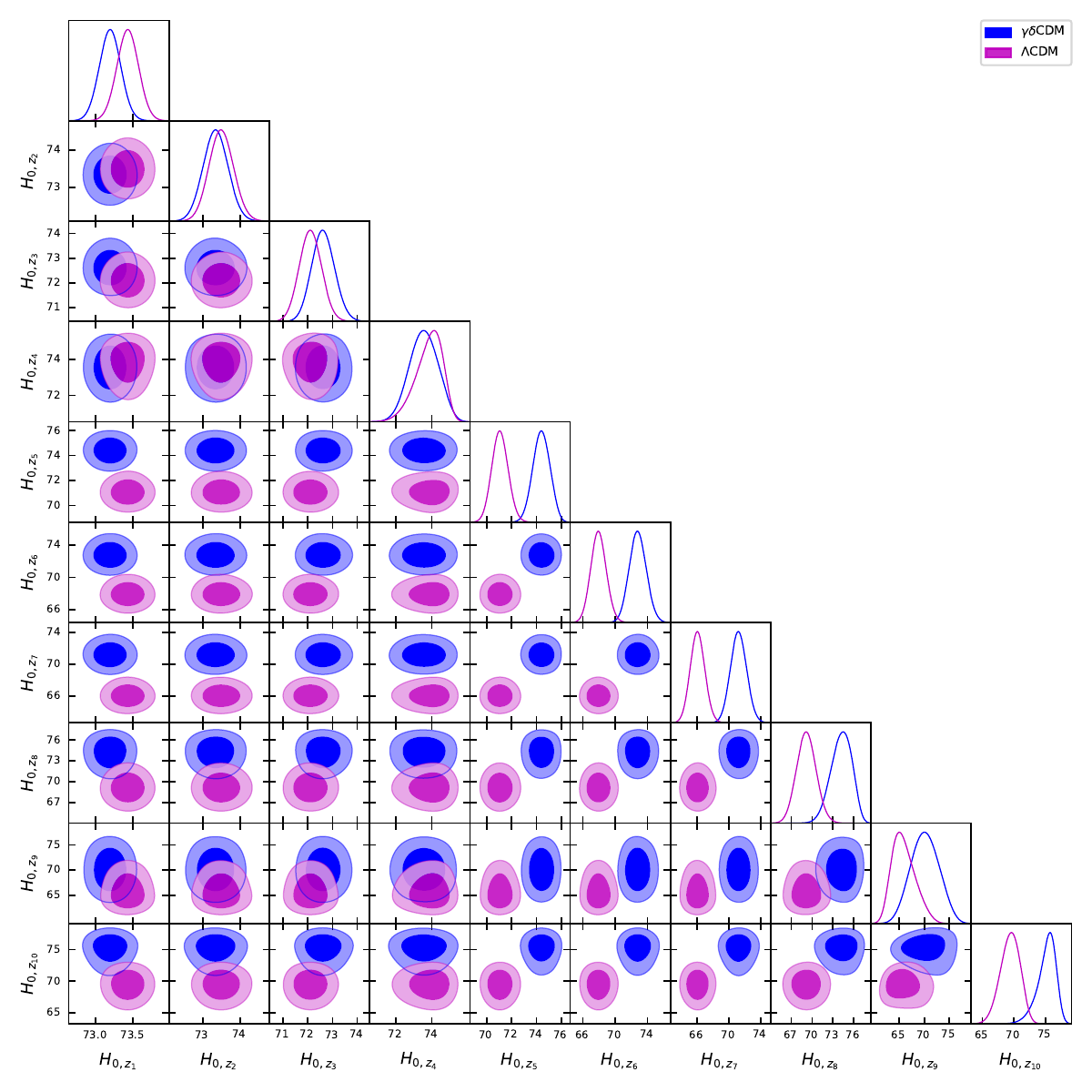}
\caption{Combined contour plots for 2D joint posterior distributions, with 1$\sigma$ and 2$\sigma$ confidence regions, for the $H_{0,z_i}$ in each equal-width bin of the CC+DESI+Pan+ dataset in the \gdcdm\ (blue) and the \lcdm\ (purple) models, together with 1D marginalized posterior distributions.}
\label{dewcp}
\end{figure}

\begin{table}[h]
\centering
\def\arraystretch{1.5}
\begin{tabular}{lccc}
\hline
\hline
Redshift bin & \vspace{1mm}\parbox[t]{15mm}{\centering Data number}
& \parbox[t]{18mm}{\centering \gdcdm\ $\mathbf{H_{0,z_i}}$}
& \parbox[t]{18mm}{\centering \lcdm\ $\mathbf{H_{0,z_i}}$} \\
\hline
$[0.0 - 0.1]$ & $632$ & $73.19 ^{+0.14}_{-0.14}$ & $ 73.43 ^{+0.14}_{-0.14}$ \\
$[0.1 - 0.2]$ & $211$ & $73.34 ^{+0.33}_{-0.33}$ & $ 73.49 ^{+0.33}_{-0.32}$ \\
$[0.2 - 0.3]$ & $263$ & $72.61 ^{+0.47}_{-0.44}$ & $ 72.10 ^{+0.44}_{-0.44}$ \\
$[0.3 - 0.4]$ & $188$ & $73.56 ^{+0.81}_{-0.80}$ & $ 73.96 ^{+0.64}_{-0.86}$ \\
$[0.4 - 0.6]$ & $189$ & $74.40 ^{+0.69}_{-0.67}$ & $ 71.06 ^{+0.65}_{-0.63}$ \\
$[0.6 - 0.8]$ & $104$ & $72.77 ^{+1.03}_{-1.02}$ & $ 67.92 ^{+0.94}_{-0.92}$ \\
$[0.8 - 1.1]$ & $16$ & $71.20 ^{+1.01}_{-0.98}$ & $ 66.03 ^{+0.91}_{-0.90}$ \\
$[1.1 - 1.5]$ & $19$ & $74.30 ^{+1.47}_{-1.60}$ & $ 69.15 ^{+1.38}_{-1.37}$ \\
$[1.5 - 2.0]$ & $9$ & $70.09 ^{+2.91}_{-2.82}$ & $ 65.53 ^{+2.72}_{-2.02}$ \\
$[2.0 - 2.4]$ & $3$ & $75.48 ^{+1.08}_{-1.67}$ & $ 69.49 ^{+1.49}_{-1.63}$ \\
\hline
\hline
\end{tabular}
\caption{$H_{0,z_i}$ posterior values in each of the equal-width bins of the CC+DESI+Pan+ dataset for the \gdcdm\ and the \lcdm\ models.}
\label{ewbmd}
\end{table}

In Fig. \ref{dewcp} we present combined contour plots for 2D joint posterior distributions, with 1$\sigma$ and 2$\sigma$ confidence regions, for $H_{0,z_i}$ in each equal-width bin of the CC+DESI+Pan+ dataset in the \gdcdm\ and the \lcdm\ models, together with 1D marginalized posterior distributions. This figure is also very convenient for distinguishing the trends of $H_{0,z_i}$ values with respect to redshift in the \gdcdm\ and the \lcdm\ models. Both the 2D and the 1D posterior distributions visually show that the constant $H_0$ paradoxically runs with redshift in the \lcdm\ model, whereas it remains a constant in the \gdcdm\ model.

\subsubsection{Comparison of the cosmological models \label{comd}} 

Now we would like to compare how well the \gdcdm\ and the \lcdm\ models are constrained by the CC+DESI+Pan+ dataset. As described in the preceding sections, we performed the Bayesian analysis with the full dataset, with equal-number bins and also with equal-width bins. We are going to use here again the information criteria we have already defined in Section \ref{comb}. In the same section, we also specified the free parameters in each model and in each binning method.

In Table \ref{icd} we present $min\ \chi^2$, AIC and BIC values for each model and also their differences, calculated for each dataset separately. The differences of the information criteria values in each of the datasets are remarkable.  Although \gdcdm\ model with constant $H_0$ is penalized for having additional parameters, the information criteria strongly favor it over the \lcdm\ model. The \gdcdm\ model with constant $H_0$ is also favored over the binned models. This is in contrast with the case presented in \cite{Jia:2024wix} for the \lcdm\ model. Binned analysis with redshift dependent $H_0$ is favored over the \lcdm\ model with constant $H_0$. Thus, we can safely conclude that the CC+DESI+Pan+ late-time observational dataset favors the \gdcdm\ cosmological model for which $H_0$ is a true constant.

\begin{table}[h!]
\centering
\def\arraystretch{1.5}
\begin{tabular}{l|lcccccc}
\hline
\hline
\parbox[c]{20mm}{Dataset} & \parbox[c]{12mm}{Model} 
& \parbox[c]{12mm}{\centering $\mathbf{min\ \chi^2}$} 
& \parbox[c]{12mm}{\centering $\mathbf{AIC}$} 
& \parbox[c]{12mm}{\centering $\mathbf{\Delta AIC}$} 
& \parbox[c]{12mm}{\centering $\mathbf{BIC}$} 
& \parbox[c]{12mm}{\centering $\mathbf{\Delta BIC}$} \\
\hline
Full dataset & \gdcdm\ & $1433.1$ & $1441.1$ & $0$ & $1462.7$ & $0$ \\
 & \lcdm\ & $1594.4$ & $1596.4$ & $155.3$ & $1601.8$ & $139.1$ \\
\hline
Equal-number bins & \gdcdm\ & $1429.4$ & $1445.4$ & $4.3$ & $1488.6$ & $25.9$\\
& \lcdm\ & $1496.7$ & $1512.7$ & $71.6$ & $1555.9$ & $93.2$ \\
\hline
Equal-width bins & \gdcdm\ & $1425.2$ & $1445.2$ & $4.1$ & $1499.2$ & $36.5$ \\
& \lcdm\ & $1485.1$ & $1505.1$ & $64.0$ & $1559.1$ & $96.4$ \\
\hline
\hline
\end{tabular}
\caption{Minimum chi-squared, Akaike Information Criterion (AIC) and Bayesian Information Criterion (BIC) results for the \gdcdm\ and the \lcdm\ models fitted to the CC+DESI+Pan+ dataset.  Differences of information criterion values are calculated with $\Delta \mathrm{AIC} = \mathrm{AIC}_{model} - \mathrm{AIC}_{\ga\de CDM}$ for each distinct model and similarly for $\Delta$BIC.}
\label{icd}
\end{table}

\section{Conclusions \label{conc}} 

The Hubble tension is not resolved yet. In contrast, it is getting worse with the new late-time observations \cite{Said:2024pwm,Scolnic:2024oth,Boubel:2024cqw,Pascale:2024qjr,Vogl:2024bum}. Related to the Hubble tension, there have been many works \cite{Wong:2019kwg,Jia:2022ycc,Millon:2019slk,Krishnan:2020obg,Dainotti:2021pqg,Schiavone:2022shz,Dainotti:2022bzg,Dainotti:2022rea,Colgain:2022rxy,Colgain:2022tql,Malekjani:2023ple,Jia:2024wix} that indicated the running of the Hubble constant with redshift in the \lcdm\ model: the value of the Hubble constant inferred from observations at different redshifts implied a redshift-dependent decreasing trend in the Hubble constant. In the $\omega_0\omega_a$CDM model \cite{Chevallier:2000qy,Linder:2002et} this decreasing trend was found even more pronounced compared to the \lcdm\ model as shown in \cite{Dainotti:2021pqg,Schiavone:2022shz,Dainotti:2022bzg}. Such a decreasing trend was also observed \cite{DeSimone:2024lvy} in Maeder's scale-free cosmology \cite{Maeder:2017ksf}. Several works \cite{Dainotti:2021pqg,Malekjani:2023ple,Schiavone:2022wvq,Montani:2024ntj,DeSimone:2024lvy} have aspired to find theoretical framework for the running of $H_0$ by modifying the \lcdm\ model. These works accepted the decay of the Hubble constant with redshift as a fact forced on us by observations and aimed to derive such a decreasing trend from the theory.
 
However, a fundamental constant should be and remain constant throughout cosmological history and thus at all redshifts. The fact that it turns out to be a function of redshift in the \lcdm\ or related models points out that there is a problem with the \lcdm\ model, indicating unknown new physics, or there are unknown systematics in some of the observations. Thus, we investigated the redshift dependence of the Hubble constant in the \gdcdm\ cosmological model to determine whether a similar behavior is also observed in this model. Through data analysis with the redshift bins of late-time cosmological data, we established that the Hubble constant in the \gdcdm\ model does not evolve with redshift. Whereas in the case of the \lcdm\ model, our results agree with the results of \cite{Jia:2022ycc} and \cite{Jia:2024wix}; in the case of the \gdcdm\ model, our results demonstrate a completely different picture than the \lcdm\ model. The value of the Hubble constant obtained in the \gdcdm\ model is in the 1$\si$ bound of the late Universe observations in almost all the redshift bins. We also confirmed that our model fits the aforementioned data better than the \lcdm\ model by checking various information criteria.
Another interesting result is that the size of the sound horizon at baryon drag in the \gdcdm\ model is calculated to be in the 1$\sigma$ range of the model-independent estimate from a recalibration of the eBOSS and the DESI BAO datasets, obtained by deep learning techniques \cite{Shah:2024gfu}. 

The \gdcdm\ cosmological model \cite{Deliduman:2023caa,Binici:2024smk} is based on $f(R)$ gravity in an anisotropic background. In this model, the expansion of the Universe depends on the energy content of the Universe very differently compared to the \lcdm\ model. The differences come about twofold:
Firstly, the contribution of each energy density to the Hubble parameter is weighted by an equation of state parameter dependent constant. Additionally, in their contribution to the Hubble parameter, the dependence of energy densities on redshift differs from what their physical nature requires. This change in the relation of the Universe's energy content to the Hubble parameter modifies the relation between redshift and the cosmic time.
The \gdcdm\ model has been successful so far. In \cite{Deliduman:2023caa} we tested this model by constraining its parameters with the low-redshift CC Hubble, the Pantheon SNe Ia, and the eBOSS BAO data together with the high-redshift CMB data. It is found that the \gdcdm\ model fits to these datasets slightly better than the \lcdm\ model, and furthermore supports the conjecture of the cosmological coupling of black holes \cite{Croker:2021duf,Farrah:2023opk,Croker:2020plg}.
We then tested \cite{Binici:2024smk} the \gdcdm\ model with the ages of the oldest astronomical objects \cite{Vagnozzi:2021tjv}, together with the CC Hubble and the Pantheon+ SNe Ia data. 
We showed that the \gdcdm\ model's modified time-redshift relation allows more time at high redshift for massive galaxies and quasars to form, though the present age of the Universe is not modified significantly.

We need to test the \gdcdm\ model with many more cosmological datasets so as to decide on its viability as a bona fide cosmological model. A further tension in the current cosmology is the $S_8$ tension \cite{Abdalla:2022yfr,Perivolaropoulos:2021jda,DiValentino:2020vvd,Nunes:2021ipq}, which might be related to the Hubble tension. Since $S_8 \propto \sqrt{\Omega_{m0}}$ \cite{Vagnozzi:2023nrq}, the value of $S_8$ increasing with redshift might point out running of $\Omega_{m0}$ with redshift. Such a trend is observed in the \lcdm\ model in the binned analysis of the quasar data \cite{Colgain:2022nlb}, of the Pantheon+ SNe data \cite{Colgain:2022rxy,Malekjani:2023ple}, and of the Dark Energy Survey SNe data \cite{Colgain:2024ksa,DES:2024jxu}. Since $\Omega_{m0}$ should also be a constant in the \lcdm\ model, its running with redshift is also very problematic. Such an increasing trend of $\Omega_{m0}$ or $S_8$ \cite{Adil:2023jtu,Akarsu:2024hsu} might be pointing out the breakdown of the \lcdm\ model or an unknown local physics, as in the case of running of the Hubble constant. We plan to analyze, with various datasets, whether the inferred values of the $\Omega_{m0}$ parameter from observations at distinct redshifts remain constant in the \gdcdm\ model.

\section*{Acknowledgments} 

Authors thank O\u{g}uzhan Ka\c{s}\i k\c{c}\i\ and Nihan Kat{\i}rc{\i} for helpful discussions. Furkan \c{S}akir Dilsiz is also supported by T\"UB\.{I}TAK-B\.{I}DEB 2211-A Domestic Ph.D. Scholarship. The numerical calculations reported in this article were partially performed at T\"UB\.{I}TAK ULAKB\.{I}M, High Performance and Grid Computing Center (TRUBA resources).



\end{document}